\documentclass[sigconf,edbt]{acmart-edbt2021}

\def\BibTeX{{\rm B\kern-.05em{\sc i\kern-.025em b}\kern-.08em
    T\kern-.1667em\lower.7ex\hbox{E}\kern-.125emX}}

\usepackage{booktabs} %

\usepackage{multirow}
\usepackage{chronosys}
\usepackage{algorithm}
\usepackage{algorithmicx}
\usepackage{algpseudocode}
\usepackage{subcaption}
\usepackage{fixltx2e} %
\usepackage{hyperref}

\usepackage[linewidth=1pt]{mdframed}

\algdef{SE}[DOWHILE]{Do}{doWhile}{\algorithmicdo}[1]{\algorithmicwhile\ #1}%

\newcommand{\bred}[1]{}
\newcommand{\iparagraph}[1]{\textbf{\textit{#1.}}}

\usepackage{amsmath}

\usepackage{pgfgantt}
\usetikzlibrary{shadows}
\usetikzlibrary{shadings}
\usepackage{float} 
\usepackage{balance}  %

\usepackage{array}
\newcolumntype{L}[1]{>{\raggedright\let\newline\\\arraybackslash\hspace{0pt}}m{#1}}
\newcolumntype{C}[1]{>{\centering\let\newline\\\arraybackslash\hspace{0pt}}m{#1}}
\newcolumntype{R}[1]{>{\raggedleft\let\newline\\\arraybackslash\hspace{0pt}}m{#1}}
\newcolumntype{M}[1]{>{\centering\arraybackslash}m{#1}}

\usepackage{enumitem}
\newlist{todolist}{itemize}{2}
\setlist[todolist]{label=$\square$}
\usepackage[colorinlistoftodos,disable]{todonotes}
\usepackage{pifont}

\setcopyright{rightsretained}

\acmDOI{}

\acmISBN{978-3-89318-084-4}

\acmConference[EDBT 2021]{24th International Conference on Extending Database Technology (EDBT)}{March 23-26, 2021}{Nicosia, Cyprus} 
\acmYear{2021}

\begin{document}
\title{Shift-Table: A Low-latency Learned Index for Range Queries using Model Correction}

\author{Ali Hadian}
\orcid{0000-0003-2010-0765}
\affiliation{%
  \institution{Imperial College London}
}

\author{Thomas Heinis}
\orcid{0000-0002-7470-2123}
\affiliation{%
  \institution{Imperial College London}
}

\renewcommand{\shortauthors}{A. Hadian and T. Heinis}

\begin{abstract}
        Indexing large-scale databases in main memory is still challenging today. Learned index structures --- in which the core components of classical indexes are replaced with machine learning models --- have recently been suggested to significantly improve performance for read-only range queries.
        
		However, a recent benchmark study shows that learned indexes only achieve limited performance improvements for real-world data on modern hardware. More specifically, a learned model cannot learn the micro-level details and fluctuations of data distributions thus resulting in poor accuracy; or it can fit to the data distribution at the cost of training a big model whose parameters cannot fit into cache. As a consequence, querying a learned index on real-world data takes a substantial number of memory lookups, thereby degrading performance. 
		
		In this paper, we adopt a different approach for modeling a data distribution that complements the model fitting approach of learned indexes. We propose \textit{Shift-Table}, an algorithmic layer that captures the micro-level data distribution and resolves the local biases of a learned model at the cost of at most one memory lookup. Our suggested model combines the low latency of lookup tables with learned indexes and enables low-latency processing of range queries. 
		Using Shift-Table, we achieve a speedup of 1.5X to 2X on real-world datasets compared to trained and tuned learned indexes. 
		
\end{abstract}

\maketitle

\section{Introduction}

Trends in new hardware play a significant role in the way we design high-performance systems. A recent technological trend is the divergence of CPU and memory latencies, which encourages decreasing random memory access at the cost of doing more compute on cache-resident data ~\cite{kraska2018case,van2019efficiently,zhang2018surf}. 

A particularly interesting family of methods exploiting the memory/CPU latency gap are learned index structures. A learned index uses machine learning instead of algorithmic data structures to learn the patterns in data distribution and exploits the trained model to carry out the operations supported by an algorithmic index, e.g., determining the location of records on physical storage~\cite{ding2020alex,galakatos2019fiting,hadian2019interpolation,kraska2019sagedb,kraska2018case,llaveshi2019accelerating}. If the learned index manages to build a model that is compact enough to fit in processor cache, then the results can ideally be fetched with a single access to main memory, hence outperforming algorithmic structures such as B-trees and hash tables.  

In particular, learned index models have shown a great potential for range queries, e.g., retrieving all records where the key is in a certain range $A < \text{key} < B$. To enable efficient retrieval of range queries, range indexes keep the records physically sorted. Therefore, retrieving the range query is equivalent to finding the first result and then sequentially scanning the records to retrieve the entire result set. Therefore, processing a range query $A < \text{key} < B$ is equivalent to finding the first result, i.e., the smallest key in the dataset that is greater than or equal to $A$ (similar to \texttt{lower\_bound(A)} in the C++ Library standard). A learned index can be built by fitting a regression model to the cumulative distribution function (CDF) of the key distribution. The learned CDF model can be used to determine the physical location where the lower-bound of the query resides, i.e., $\texttt{pos}(\text{A}) = N\times F_\theta(\text{A})$ where N is the number of keys and $F_\theta$ is the learned CDF model with model parameters $\theta$.   

Learned indexes are very efficient for sequence-like data (e.g., machine-generated IDs), as well as synthetic data sampled from statistical distributions. However, a recent study using the Search-On-Sorted-Data benchmark (SOSD)~\cite{kipf2019sosd} shows that for real-world data distributions, a learned index has the same or even poorer performance compared to algorithmic indexes. For many real-world data distributions, the CDF is too complex to be learned efficiently by a small cache-resident model. The data distribution of real-world data has "too much information" to be accurately represented  by a small machine-learning model, while an accurate model is needed for an accurate prediction. One can of course use smaller models that fit in memory with the cost of lower prediction accuracy, but will end up in searching a larger set of records to find the actual result which consequently increases memory lookups and degrades performance. Alternatively, a high accuracy can be achieved by training a bigger model, but accessing the model parameters incurs multiple cache misses and also increases memory lookups, reducing the margins for performance improvement. 

\todo[inline]{Explain that while a lookup table (like hash table) is efficient for retrieving point queries with a single lookup, it cannot be used for range queries (WHERE X <= Q). Our solution combines the benefits of hash table (low-latency lookup) and range indexes}

In this paper, we address the challenge of using learned models on real-world data and illustrate how the micro-level details (e.g., local variance) of a cumulative distribution can dramatically affect the performance of a range index. We also argue that a pure machine learning approach cannot shoulder the burden of learning the fine-grained details of an empirical data distribution and demonstrate that not much improvement can be achieved by tuning the complexity or size thresholds of the models. 
\todo[inline]{Illustrate with a chart $\Uparrow$}

We suggest that by going beyond mere machine learning models, the performance of a learned index architecture can be significantly improved using a complementary enhancement layer rather than over-emphasizing on the machine learning tasks. Our suggested layer, called \textit{Shift-Table} is an algorithmic solution that improves the precision of a learned model and effectively accelerates the search performance. Shift-Table, targets the micro-level bias of the model and significantly improves the accuracy, at the cost of only one memory lookup. The suggested layer is optional and applied after the prediction; it can hence be switched on or off without re-training the model. 

Our contributions can be summarized as follows: 
\begin{itemize}
\vspace{-5pt}
    \item We identify the problem of learning a range index for real-world data, and illustrate the difficulty of learning from this data.
    \item We suggest the Shift-Table approach for correcting a learned index model, which complements a valid (monotonically increasing) CDF model by correcting its error.
    \item We show how, and in which circumstances, the suggested methods can be used for best performance. 
    \item We suggest cost models that determine whether the Shift-Table layer can boost performance.
    \item The experimental results show that our suggested method can improve existing learned index structures and bring stable and almost-constant lookup time for real-world data distributions. Our enhancement layer achieves up to 3X performance improvement over existing learned indexes. More interestingly, we show that for non-skewed distributions, the Shift-Table layer is effective enough to help a dummy linear model outperform the state of the art learned indexes on real-world datasets
\end{itemize}

\section{Motivation}
\label{sec:motivation}
\subsection{Lookup Cost for Learned Models}
\label{subsec:cost-of-local-search}
In modern hardware, the lookup times of in-memory range indexes and the binary search algorithm are mainly affected by their memory access pattern, most notably by how the algorithm uses the cache and the Last-Level-Cache (LLC) miss rate. %

\begin{figure*}
  \begin{minipage}[b]{.95\textwidth}
    \centering
    \includegraphics[width=.7\linewidth]{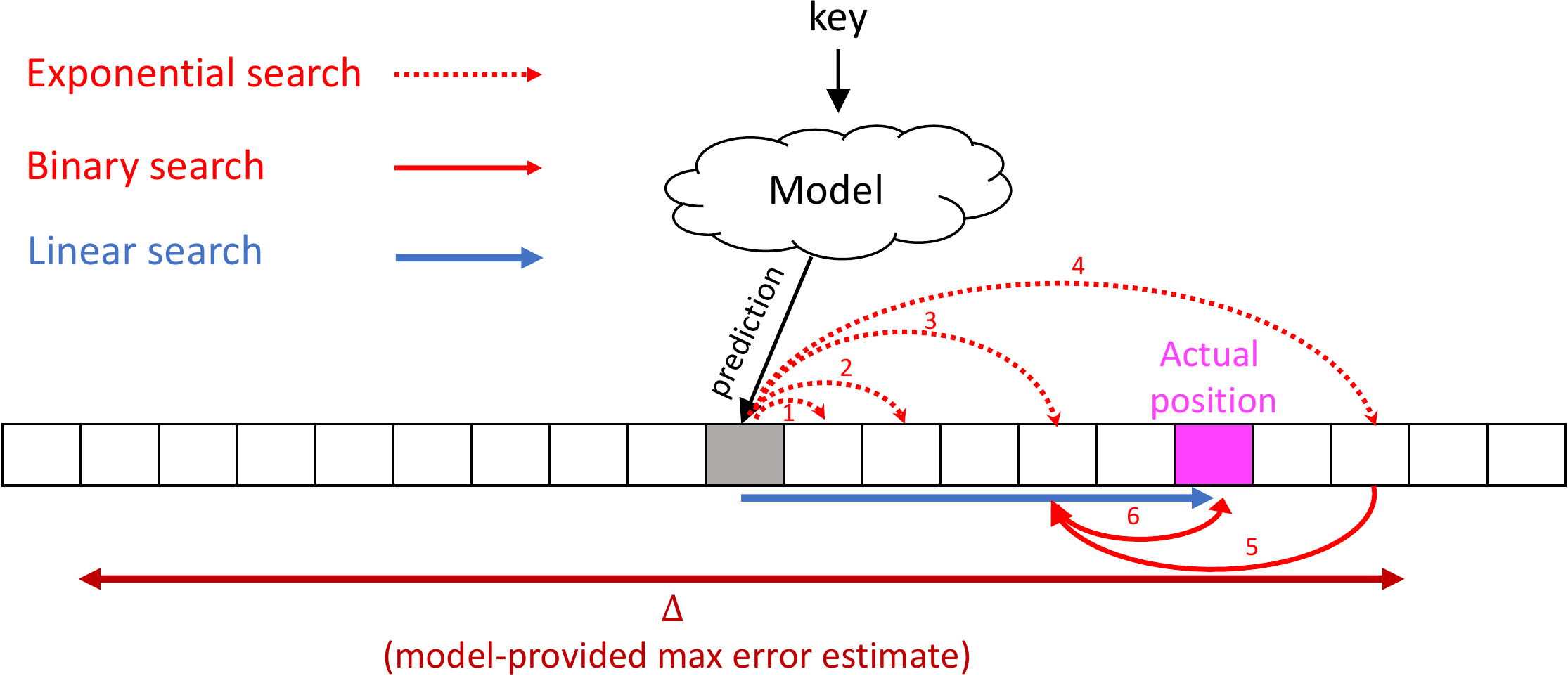}
    \subcaption{Different "last-mile" search methods (performed after location prediction) in learned index. The locations predicted by the model depend on the query and are not known in advance. Since the last-mile search algorithms need to access different memory locations for each query, they cannot exploit the processor cache and the search algorithm incurs multiple cache misses}
    \label{fig:local_search_algorithms_in_learned_index}
     \vspace{10pt}
  \end{minipage}
  \hfill
  \begin{minipage}[b]{.95\textwidth}
    \centering
    \includegraphics[width=.85\linewidth]{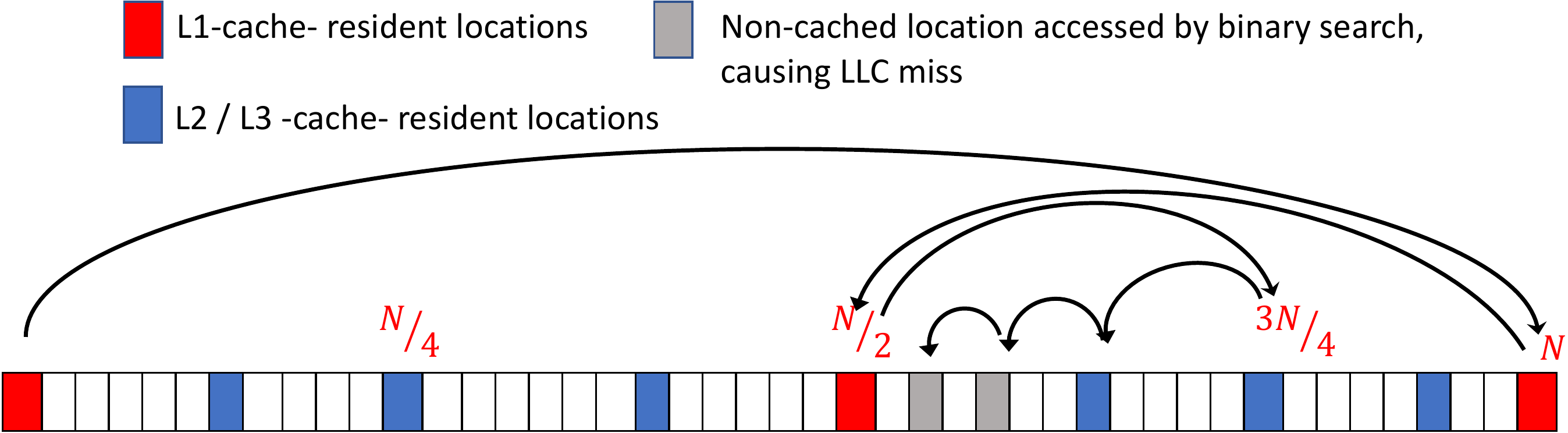}
    \subcaption{Schematic illustration of processor caching in binary search. The locations accessed by the very early stage of binary search, such as the min, max, and the midpoint, are frequently accessed and available in the L1 cache. Further steps of binary access locations that are less frequently accessed and fit on lower levels of the memory hierarchy. Therefore, a deterministic search algorithm like binary search enjoys a high cache hit rate}
    \label{fig:caching_in_binary_search}
  \end{minipage}
    \caption{Comparison of patterns in binary search (partially cached) and local search in learned indices (non-cached). }
    \label{fig:learned_index_vs_binary_search}
\end{figure*}

Processing a range query in a learned index has two stages: 1)~\textbf{Prediction}: Running the learned model to predict the location of the first result for the range query; and 2)~\textbf{Local search} (also known as \textit{last-mile search}): searching around the predicted location to find the actual location of the first result. Figure~\ref{fig:local_search_algorithms_in_learned_index} shows common search methods for the local search. If the learned model can determine a guaranteed range area around the predicted position, one can perform binary search. Otherwise, exponential or linear search should be used, starting from the predicted position.

A cache miss in a learned index can occur in the first stage for accessing the parameters of the model (if the model is too big to fit in cache), or in stage two for the local search. Key in understanding the cost of a learned index is that local search is done entirely over non-cached blocks of memory. A learned index built over millions of records could predict the location of records with an error of, say, 1000 records and yet achieve no performance gain over binary search algorithms or algorithmic indexes. This is because while the learned index fits the models in cache, its algorithmic competitors also fit the frequently-accessed parts of the data in cache, which limits the potential for improvement for a learned index.

\subsection{Lookup Cost for Algorithmic Indexes}
Classical algorithms, such as binary search, can be seen as a hierarchy of [non-learned] models, which take the middle-point as its parameter and predicts (accurately) which direction the search should follow. Specifically for the first few steps of binary search where the middle-points usually reside in cache, the functionality of binary search is the same as a learned model from a performance point of view. %

In a pure binary search on the entire data, the first set of memory locations accessed by the algorithm (i.e., the median, quarters, etc.) will already be in the CPU cache after a few lookups. Therefore, the major bottleneck in binary search is for the latter stages of search where the middle elements are not in cache, causing last-level-cache (LLC) misses. Figure~\ref{fig:caching_in_binary_search} shows a schematic illustration of how caching accelerates binary search.

In basic implementations of binary search, the ``hot keys'' are cached with their payload and nearby records in the same cache line, which wastes cache space. Binary search thus uses the cache poorly and there are more efficient algorithmic approaches whose performance is not sensitive to data distributions. 

Cache-optimized versions of binary search, e.g., a binary search tree such as FAST~\cite{kim2010fast}, a read-only search tree that co-locates the hot keys but still follows the simple bisecting method of binary search, are up to 3X faster than binary search~\cite{kipf2019sosd}. This is because FAST keeps more hot keys in the cache and hence it needs to scan a shorter range of records in the local search phase (cache-non-resident iterations of the search).

\subsection{Preliminary Experimental Analysis}
\label{subsec:Preliminary-Experimental-Analysis}
\begin{figure}
  \begin{minipage}[b]{.4\textwidth}
    \centering
    \includegraphics[width=\linewidth]{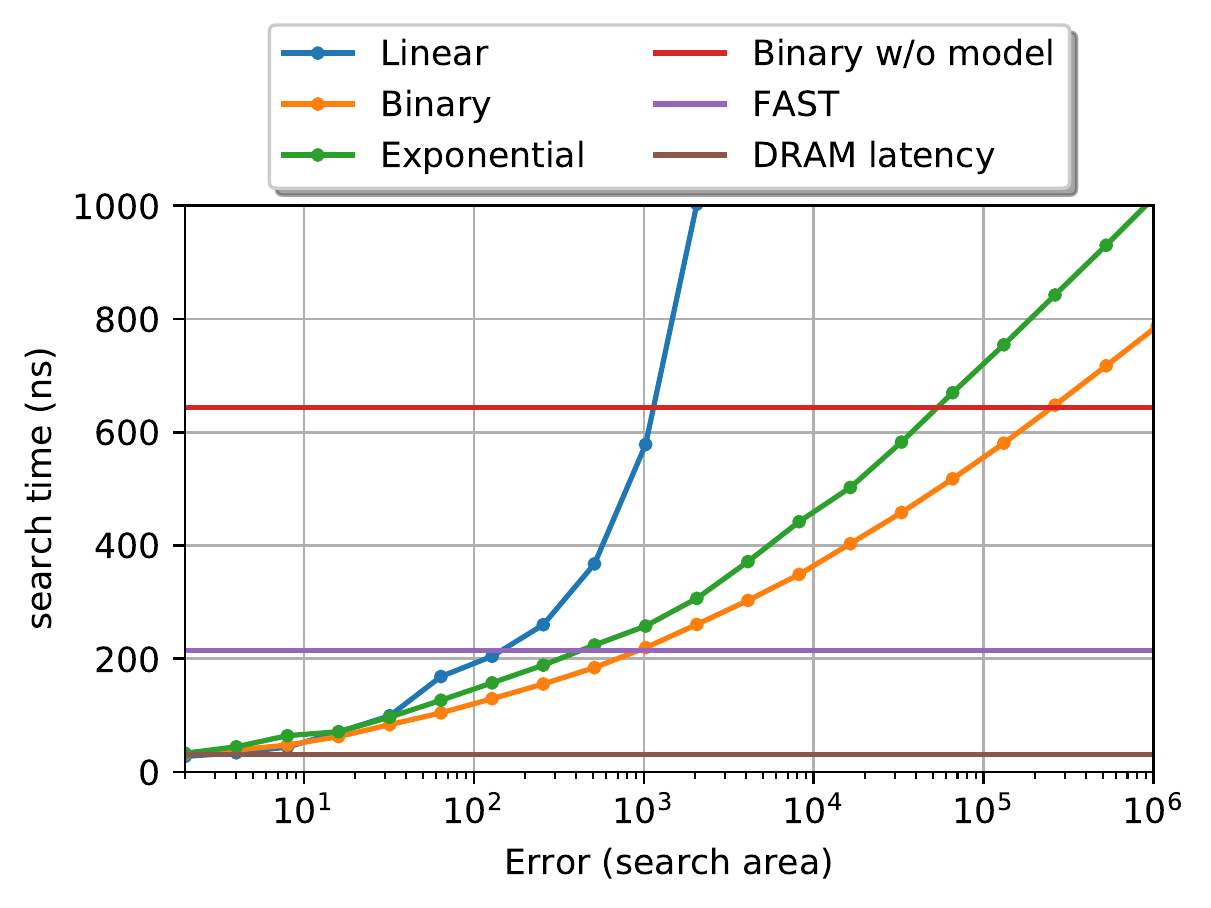}
    \subcaption{Lookup time}
    \label{fig:binary_search_vs_local_search_time}
  \end{minipage}
  \hfill
  \begin{minipage}[b]{.4\textwidth}
    \centering
    \includegraphics[width=\linewidth]{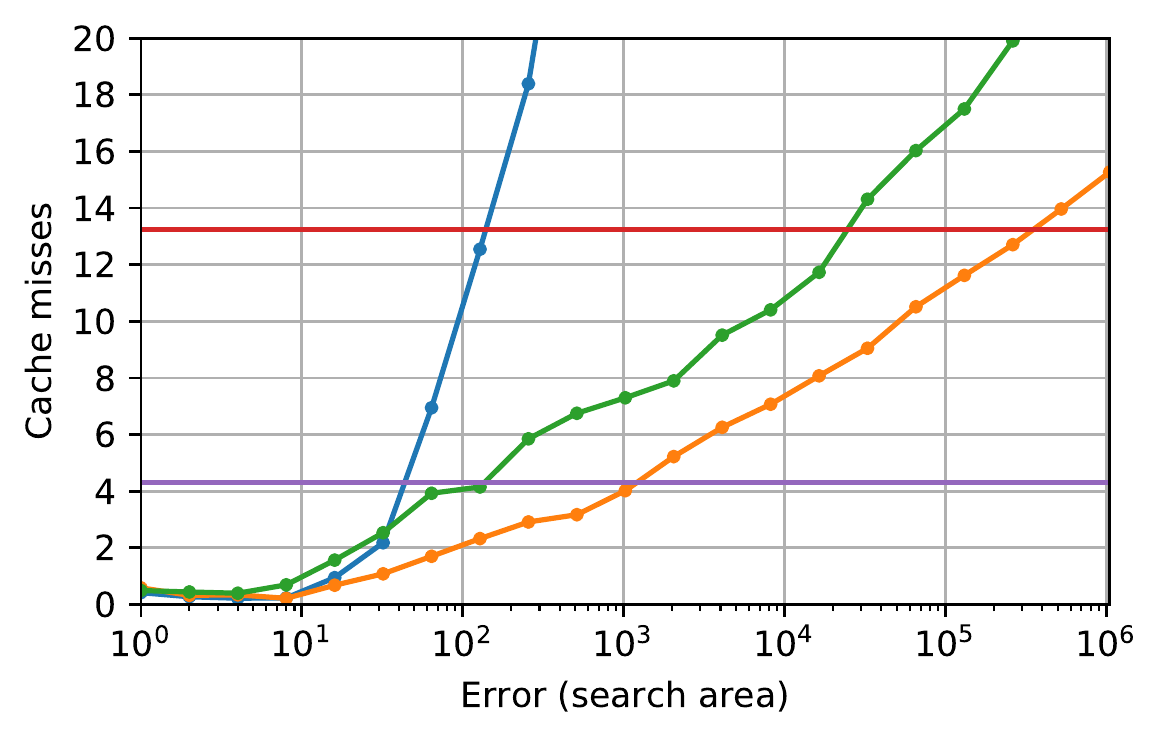}
    \subcaption{Cache misses}
    \label{fig:binary_search_vs_local_search_LLCMiss}
  \end{minipage}
  \hfill
  \caption{Cost of local search in a learned index}
  \label{fig:search_performance_in_learned_index}
  \vspace{-10pt}
\end{figure}

For a tangible discussion and to elaborate on the real cost of a learned model, we provide a micro-benchmark that measures the cost of errors in a learned index. We use the experimental configuration used in the SOSD benchmark~\cite{kipf2019sosd}, i.e., searching over 200M records with 32-bit keys and 64-bit payloads. Figure~\ref{fig:binary_search_vs_local_search_time} shows the lookup time of the second phase (local search) in a learned model for different prediction errors. We include the lookup times for binary search, as well as FAST~\cite{kim2010fast}, over the whole array of 200M keys. 

We are interested to see that if the  position predicted by a learned index, say $\texttt{predicted\_pos}(x)$, has an error $\Delta$, then how long does it take in the local phase to find the correct record. Thus, for each query $x_i$, we pre-compute the `output' of the learned index with error $\Delta$, i.e., $[\texttt{predicted\_pos}(x_i) \pm \Delta ]$, and then run the benchmark given $\lbrace x_i,  [\texttt{predicted\_pos}(x_i) \pm \Delta ] \rbrace $ tuples.  

As shown in Figure~\ref{fig:binary_search_vs_local_search_time}, if the error of the model is more than $\sim$300 records on average, then FAST outperforms the learned model (with linear or exponential local search). Even if the learned model can give a guaranteed range around the predicted point to guide the local search and enable binary search, FAST outperforms it if the error exceeds 1000 records. The same trend can be seen for the LLC miss rates in Figure~\ref{fig:binary_search_vs_local_search_LLCMiss}.

Note that this micro-benchmark over-estimates the maximum error that the learned index can have because we only compare the time of \textit{local search phase} in a learned index with the \textit{total search time} of FAST and binary search. Considering the time taken to execute the model for predicting the location, a learned model needs to have a much lower error to compete with the generic, reliable, and distribution-independent algorithms such as binary search and FAST. For example, FAST takes 200 nanoseconds to search a key in the entire 200M-key dataset. If a learned index takes, say, 120 nanoseconds to run (for accessing model parameters and computing the prediction),  then the local search can take at most 80 nanoseconds so that the learned index can outperform FAST, which means that the prediction error ($\Delta$) must be less than 16 records (based on Figure~\ref{fig:binary_search_vs_local_search_time}).

Tuning the learned index for a balance of model size and accuracy is a challenging task. Improving the local search time requires using a more accurate model with a higher learning capacity and more parameters. However, accessing such a big model typically incurs further cache misses during model execution, and consequently the lookup time. Therefore, if the data distribution cannot be learned efficiently with a small memory footprint (fitting into cache), outperforming cache-efficient algorithmic indexes is very challenging. This is indeed the case for most real-world datasets that cannot be modelled accurately with a small-sized model.

\subsection{Difficulty of Learning Real-world Data}
\label{subsec:motivation-difficulty}
To use a learned index in a production system, it is essential to identify when learned indexes fail to achieve superior performance and what aspects of the data distribution contributes to the performance of a learned index model. We realized that a major challenge in understanding learned indexes is that the common practices of performance evaluation for indexing algorithms are misleading for learned indexes. For example, it is common to use the uniform and skewed distributions (such as log-normal) as arguably the two best- and worst-case extremes for a search task~\cite{kraska2018case}. However, for evaluating search over sorted read-only data, the difficulty of the task is determined by the \textit{unpredictability} of the data, which is not necessarily a factor of skewness or shape parameter of the data distribution. As we will show in this section, most statistical distributions are much easier to model than real-world data.

\todo[inline]{The message of the following paragraph (synthetic distributions are not a good test case for learned indexes) is now discussed in the newer SOSD paper. Better if we mention that if we are going to cite the new SOSD paper}

\begin{figure}
  \begin{minipage}[b]{.2\textwidth}
    \centering
    \includegraphics[width=\linewidth]{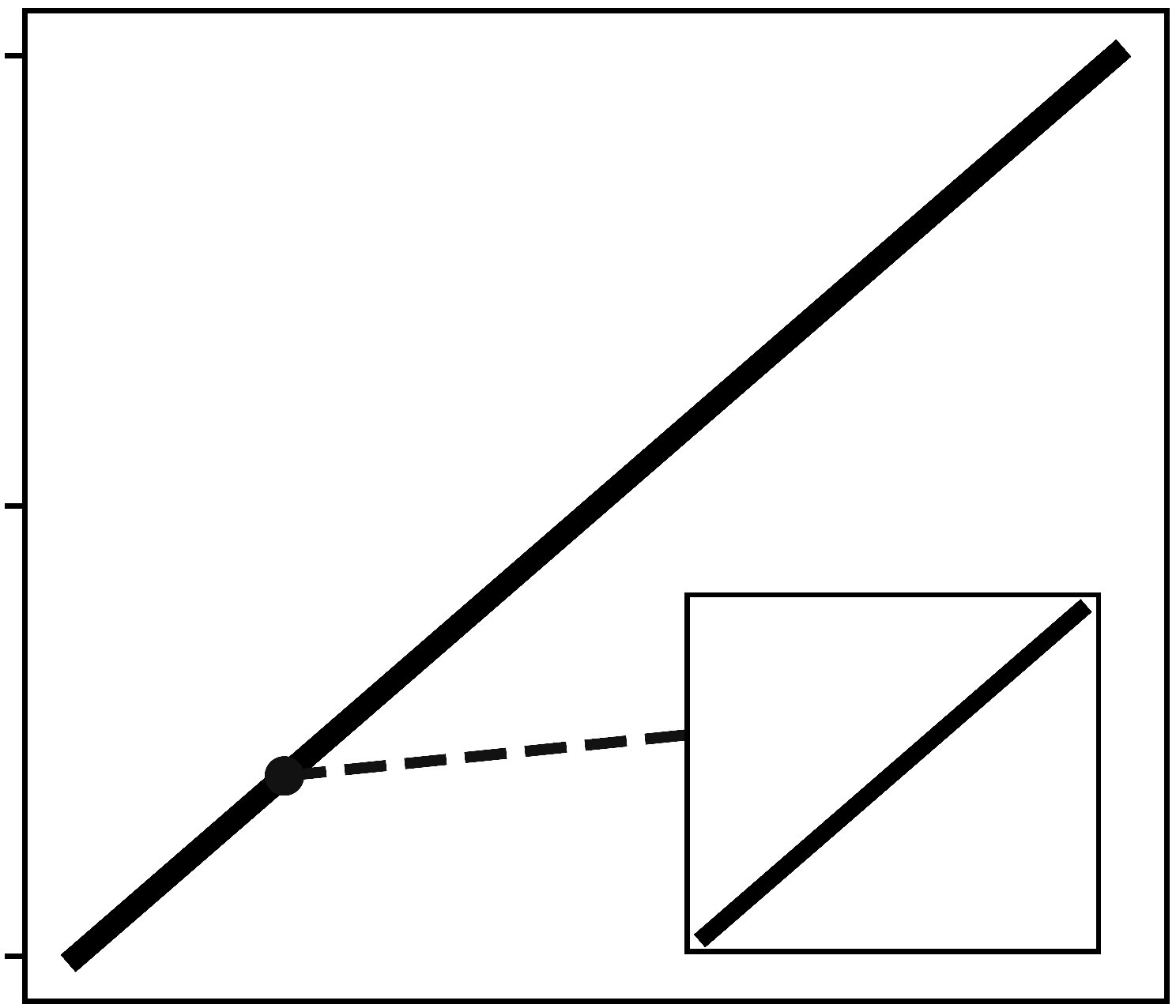}
    \subcaption{uniform}
    \label{fig:uden}
  \end{minipage}
  \hspace{10pt}
  \begin{minipage}[b]{.2\textwidth}
    \centering
    \includegraphics[width=\linewidth]{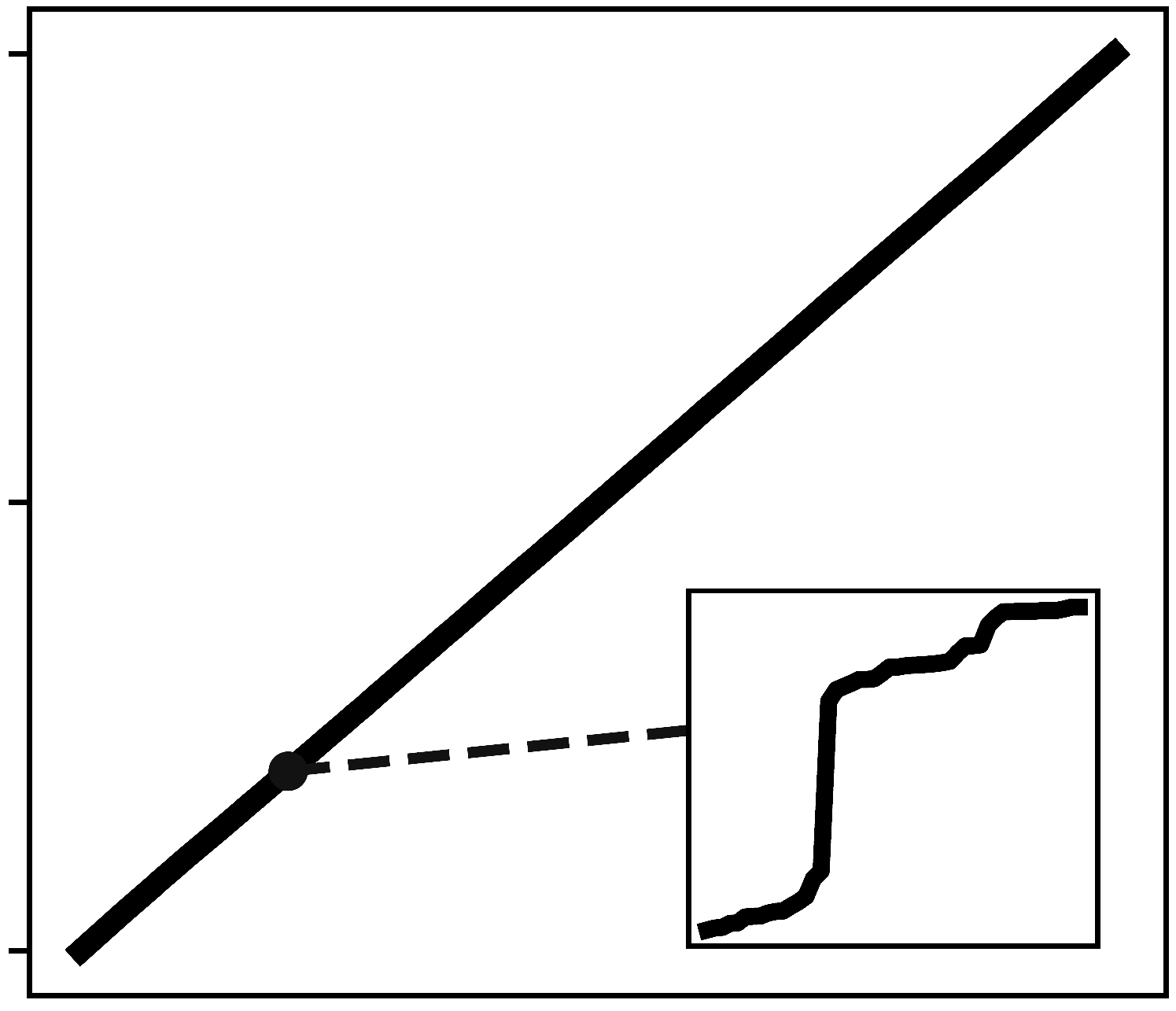}
    \subcaption{Facebook}
    \label{fig:face}
  \end{minipage}
  \begin{minipage}[b]{.2\textwidth}
    \centering
    \includegraphics[width=\linewidth]{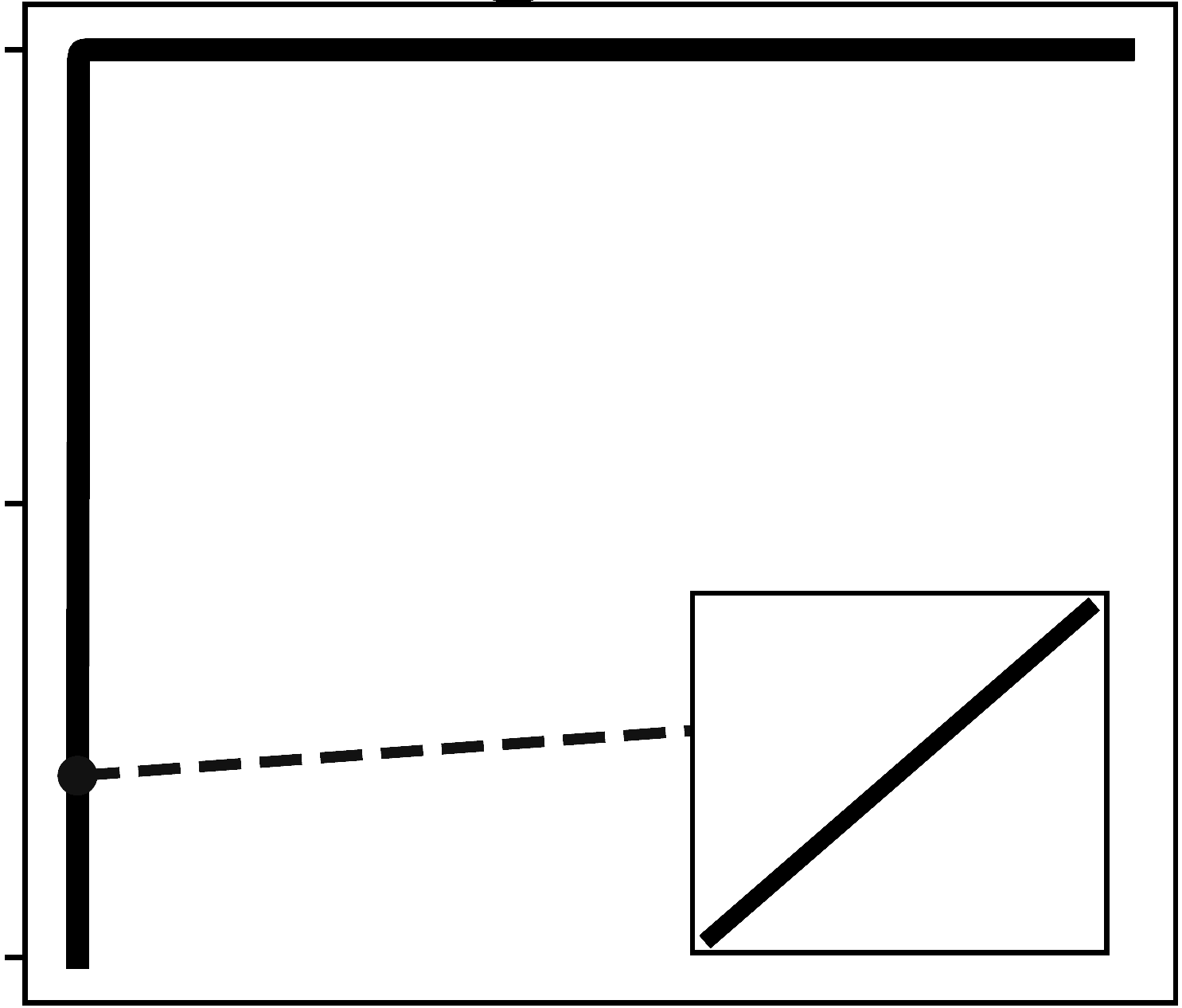}
    \subcaption{Lognormal}
    \label{fig:lognormal}
  \end{minipage}
  \hspace{10pt}
  \begin{minipage}[b]{.2\textwidth}
    \centering
    \includegraphics[width=\linewidth]{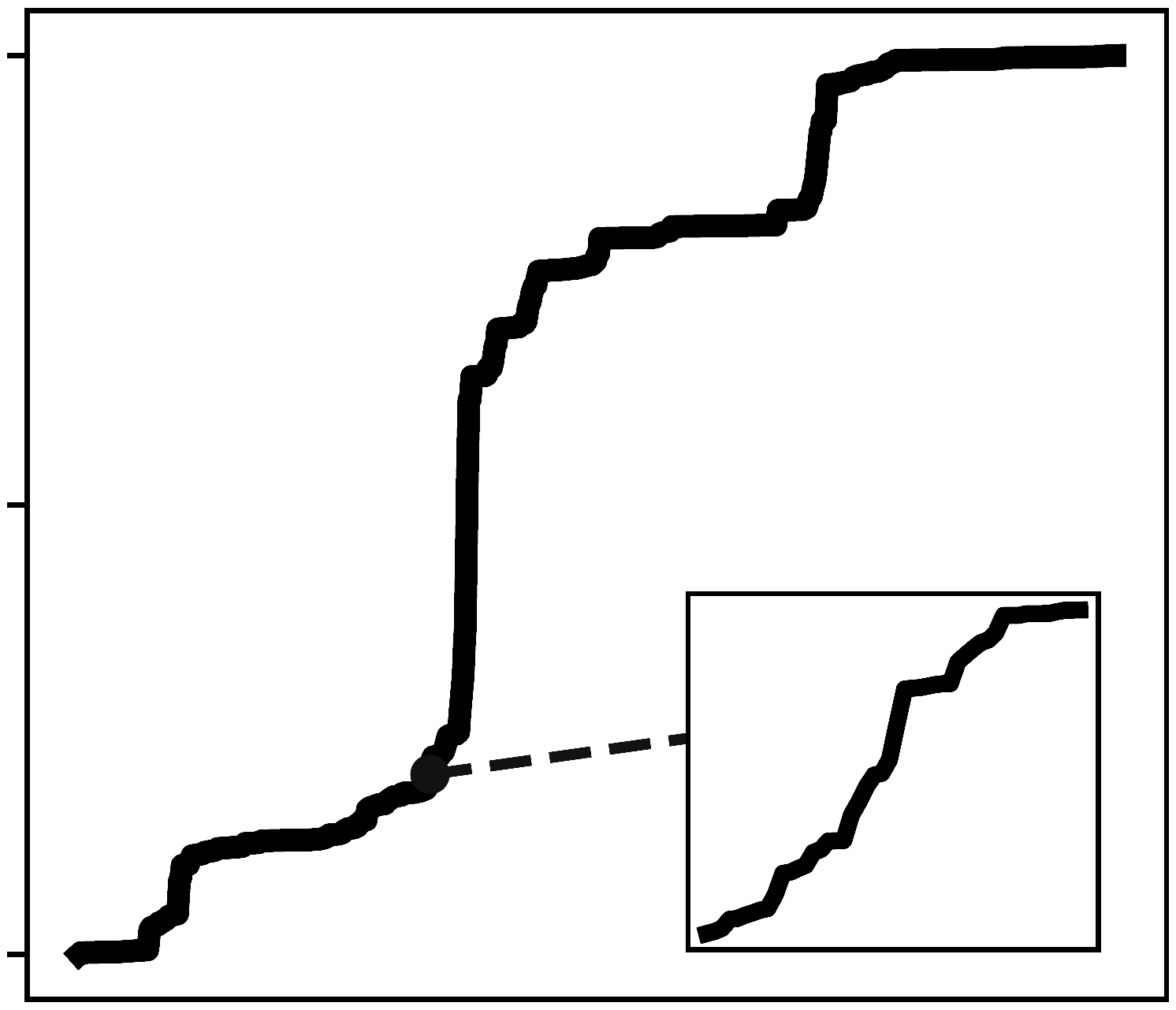}
    \subcaption{OSMC}
    \label{fig:osmc}
  \end{minipage}
  \caption{Example distributions with different complexities in micro and macro levels}
  \label{fig:example-distributions}
\end{figure}

\paragraph{Distributions that matter}
An interesting observation from the SOSD benchmark results is that even for datasets that have the same background distribution, e.g., both closely match a uniform distribution, the performance of a learned model can vary significantly, depending on the fine-grained details in the empirical CDFs. For example, consider Figures~\ref{fig:uden} and \ref{fig:face}, which represent two CDFs that are both close to uniform. The uniform data (\texttt{uden64} ~\cite{kipf2019sosd}) is comprised of dense integers that are synthetically sampled from a uniform distribution, and Facebook (\texttt{face64} \cite{kipf2019sosd}) is a Facebook user ID dataset. While both datasets match closely with the uniform distribution, face64 is significantly harder to model due to its fine-grained details in the CDF. The lookup time of learned indexes (both RMI and Radix-Splines) for face64 is 6-7$\times$ higher than that of uden64 (see Table~\ref{tab:sosd_results}) because there are many micro-level details (unpredictability) in the CDF, hence a huge model with a high learning capacity is needed to fit the CDF accurately. Using the RMI learned index, for example, the uden64 data is easily modelled with a simple line (two parameters) with near-zero error, while the best architecture found by the SOSD benchmark for modelling the face64 data is a hierarchy of two linear models, a huge model (136MB), with an average error of 202 records.

Generally speaking, real-world datasets are more difficult to learn compared to synthetic ones and the learned index built over them is not significantly faster than the algorithmic rivals. The main question remains what distinguishes a real-world data from a synthetic one? Consider the four distributions in Figure~\ref{fig:example-distributions}, where Figures~\ref{fig:uden}, \ref{fig:lognormal} are synthetic (generated from uniform and log-normal distributions), and Figures~\ref{fig:face}, \ref{fig:osmc} are real-world data. 
 The mini-chart inside each CDF highlights the distribution in a small sub-range, i.e., a ``zoomed-in'' view of the CDF. For the synthetic data, the CDF is very smooth in any short sub-range of the whole CDF. Synthetic data (such as uniform, normal and log-normal) are built using a cumulative density function that is derivable, meaning that the at any small sub-range, the shape of the CDF is close to a straight line with a slope that is close to the derivation of the underlying CDF in that range. Such a smooth CDF has less information to be compressed into a model. For example, a learned index model based on linear splines can accurately fit the whole CDF by fitting each part of the CDF to a line. Even for very skewed distributions, such as log-normal, the data is so predictable that it can be easily fitted to simple, linear models. 

Real-world data, however, is much less predictable and has a much higher level of complexity in its patterns. Even if an ideal learning algorithm is used to model the real-world data, the model itself needs to be very big because the compressed version of the CDF (to be stored as a model) is still very big.

This explains why state-of-the-art learned indexes perform extremely well for datasets that are synthetically generated from a statistical distribution (such as uniform, normal, and log-normal), but perform comparably poor for real-world data that even almost match (shapewise) with those synthetic distributions~\cite{kipf2019sosd}. On real-world datasets, learned indexes have a high cache miss rate and lookup time, contrary to their primary goal of having fewer cache misses. %

Using learned models is beneficial when they are 1) accurate enough to predict a position within the same cache line that contains the data point, otherwise the lookup time will be adversely affected due to multiple cache misses, and 2) compact enough to fit in cache and not to cause LLC misses. With this in mind, we can argue that a pure machine-learning approach might fail to ``learn the data perfectly'' and ``fit the model in cache'' simultaneously, specifically in case of real-world datasets that contain a lot of underlying patterns like spikes and generally noise. 

As a consequence, learned models are crucial to indexing but they cannot shoulder the burden of indexing the data alone. We hence suggest an algorithmic layer that can mitigate the difficulty of learning the data distribution. In this approach, the learned model is allowed to learn an semi-accurate, small model that learns the holistic shape of the distribution, and the fine-tuned modelling is provided by the algorithmic layers.

\todo[inline]{Now that we have only one layer (Shift-Table), this subsection seems to be too verbose.}
\subsection{Model Correction}

While learned index models are powerful tools for describing a data distribution in a compact representation, merely focusing on learning a highly-accurate model does not necessarily lead to a high-performance index. In this paper, we suggest a new approach for boosting existing learned models with additional layers, specifically developed with hardware costs in mind. 

The suggested helping layers add a small overhead when executing queries, but significantly reduce the overall lookup time of the learned index. The suggested layers are very powerful and consequently allow for using more lightweight models, yet ideally avoid computationally-expensive algorithms for training.

\begin{figure}
  \begin{minipage}[b]{.22\textwidth}
    \centering
    \includegraphics[width=\linewidth]{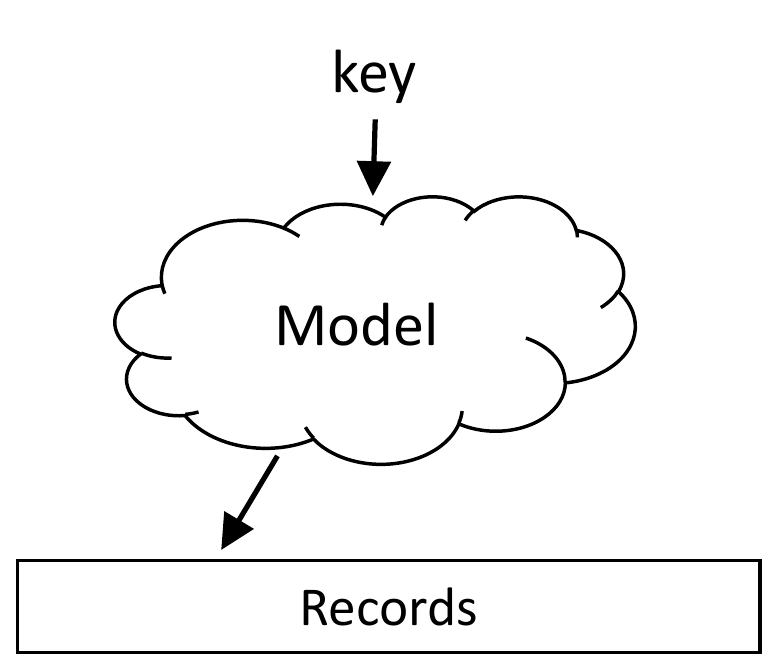}
    \subcaption{Learned index}
    \label{fig:schema_learned_index}
  \end{minipage}
  \hfill
  \begin{minipage}[b]{.22\textwidth}
    \centering
    \includegraphics[width=\linewidth]{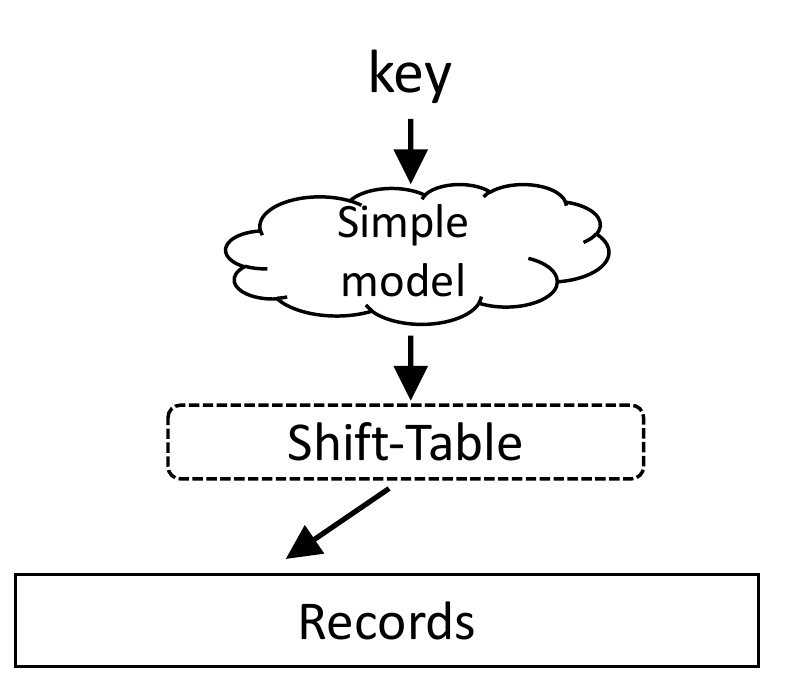}
    \subcaption{Model + Shift-Table}
    \label{fig:schema_learned_index_with_correction}
  \end{minipage}
    \caption{Leveraging correction layers to a learned index}
    \label{fig:learned_index_vs_our_solution}
    \vspace{-10pt}
\end{figure}

As Figure~\ref{fig:learned_index_vs_our_solution} illustrates, in addition to the learned index model we add a correction layer, an optional component, that can be added to improve the performance. We explore the potential of correction layers in the next sections.

\section{Shift-Table}
\label{sec:direct-mapping}
A learned model predicts a relative position $F_\theta(x)$ for a given query $x$. To calculate the position of the result, the estimated relative position is multiplied by the number of keys, and truncated to an integer (the index), hence the predicted position is $[N F_\theta(x)]$. The actual position of the record, however, is $NF(x)$ where $F(x)$ is the empirical CDF of the data points, and $N$ is the data size. Therefore, the result is $NF(x) - [N F_\theta(x)]$ records ahead of the predicted position. We identify $N F(x) - [N F_\theta(x)]$ as the \textit{drift} of $F_\theta$ at key $x$, which is the signed error of the prediction, as opposed to the absolute error. 

The idea of the Shift-Table layer is to have a lookup table that contains the drift values so that the drift of the prediction can be corrected. Capturing the drift for every value of $x$ requires an auxiliary index, which is not feasible. However, we can use the \textit{output} of the learned index model ($[N F_\theta(x)]$), which is in the range of $[0,N]$, and we construct a mapping from each possible output of the model, say $k$, to ``how far ahead is the actual record if model predicts k's record'', so that we can correct the predictions using this mapping. This means that for each prediction, we only need an extra lookup of $k$ in a fixed array of size $N$.

\todo[inline]{Explain that this is very common nowadays to have systems with lots of [extra] memory, but they are all bound to the relatively high memory latency. Therefore, it is a desirable solution for many applications to have extra O(N) space to substantially decrease latency.}

To build the Shift-Table layer, we first partition the keys $x_0,\cdots,x_{N-1}$ into $N$ partitions. We define $P_k$ as the set of keys for which the model predicts $k$ as the position:
\begin{equation}
    P_k = \{x \; | \; [N F_\theta(x)] = k\}
\end{equation}

Each of the indexed keys in $P_k$ has an index, say $NF(x)$ and a prediction $k=[N F_\theta(x)]$. 
For each partition, we extract two parameters that specify the range for local search, namely $\Delta_k$ and $C_k$.  $\Delta_k$ is defined as:
\begin{equation}
    \Delta_k = \text{min}\left(N F(x) - k]\right)\;\; \forall x \in P_k
\end{equation}

which indicates that if the predicted location is $k$, the search should be started at point $k + \Delta_k$. Also, $C_k = \vert P_k \vert$ is the cardinality of $P_k$, i.e., the number of indexed keys for which the prediction predicts the $k$'th record, in other words, the length of the area that has to be searched in the local search phase. 

To correct the prediction, we first compute the predicted position $k=[N F_\theta(x)]$, and then perform local search in the range of $[k+\Delta_k , k+\Delta_k+C_K-1]$. 

The number of partitions depends on the range of the output of the learned index, which should be $0,N)$. Therefore, The  $\texttt{<}\Delta_k,C_k\texttt{>}$ pairs: pairs are stored in a single array of size $N$, so that the correction can be done using a single lookup into the array of pairs. 

\begin{figure*}
    \centering
    \includegraphics[width=.95\linewidth]{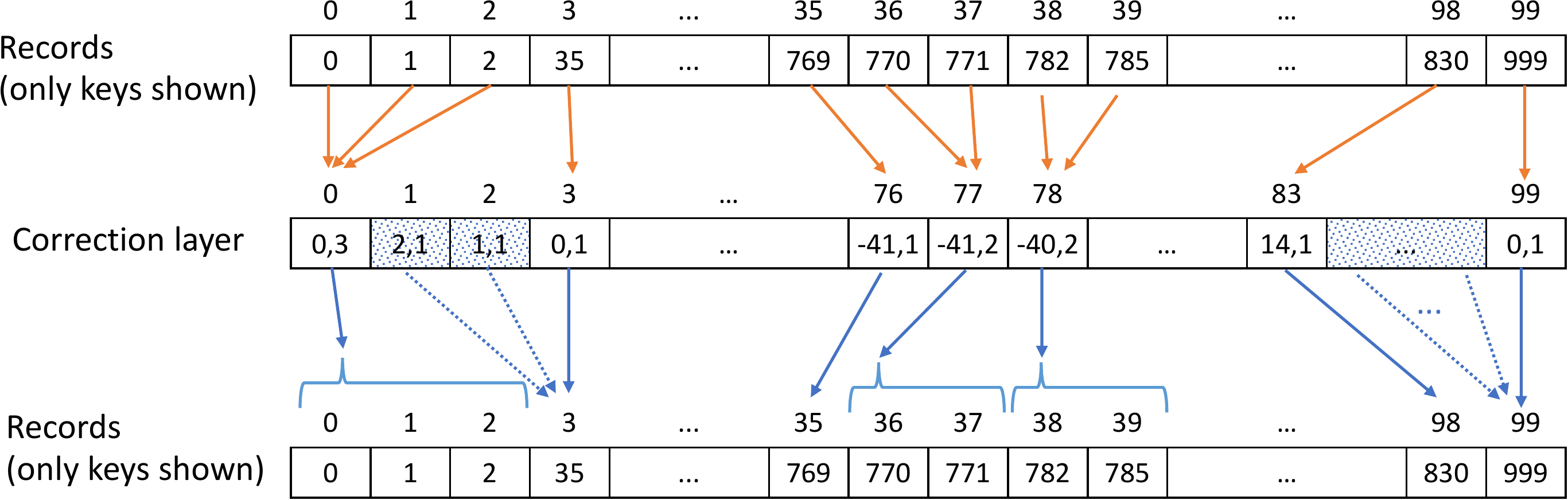}
    \caption{Shift-Table}
    \label{fig:direct-mapping}
  \end{figure*}

A Shift-Table layer is depicted in Figure~\ref{fig:direct-mapping}. The index contains 100 elements in range [0,999]. The CDF model is a simple model: $F_\theta(x) = x/1000$, hence the prediction is simply $k=[x/10]$. If the query is 771, for example, the prediction of the model is $k=77$. The correction information are $\Delta_{77}=-41$ and $C_{77}=2$, which indicates that the result is -41 records ahead of the prediction, and the search area is of length 2. Therefore, the local search is performed on the indexes of range $[36,37]$.

Algorithm~\ref{alg:search} shows how Shift-Table is used to accelerate query processing. The Shift-Table layer reduces the prediction error of the model, but incurs an additional memory lookup.

\begin{algorithm}
    \caption{Search with direct-mapped learned index}
    \label{alg:search}
    \begin{algorithmic}[1] %
        \Procedure{FIND\_LOWER}{$q$, model, Shift-Table} 
        \State pos = model.predict(q)
        \State pos = Shift\_Table.mapping[pos].startPoint %
        \State range = Shift\_Table.mapping[pos].range
        \If{range $<$ linear\_to\_binary\_threshold}
            \State pos = LinearSearch(start=data[pos],range)
        \Else
            \State pos = BinarySearch(start=data[pos],range)
        \EndIf    
        \State \Return pos
        \EndProcedure
    \end{algorithmic}
\end{algorithm}

\subsection{Querying non-indexed keys}
If the query is on the indexed keys, the result is in range $[k+\Delta_k \; , \; k+\Delta_k+C_K-1]$. In  Figure~\ref{fig:direct-mapping}, for example, querying 771 and 782 points to the correct range that contains the result. However, if the query is not among the indexed keys, then the query is either within the range, or in the position just after the range (at $\text{data}[k+\Delta_k+C_K]$. For example, in Figure~\ref{fig:direct-mapping}, the record corresponding to queries 778 and 781 is the same, though the aforementioned model ($k=[q / 10]$), maps 778 to range $[36,37]$, and 781 to $[38,39]$. In both cases, however, the local search algorithm (either binary or linear search) within the range computes the correct position of the result (i.e., 38). Notably for $q=778$, a typical local search implementation realizes that the query is greater than the largest value in range and returns the first index right after the range of $[36,37]$, which is 38.

Another issue that can arise for non-indexed keys is when the predicted position $P_k$ has an empty partition that none of the indexed keys belongs to. In Figure~\ref{fig:direct-mapping}, if the query is 15, then the predicted position is $k=[15/10]=1$, but $P_1$ is empty because the model does not predict position 1 for any of the indexed keys. If the query is predicted to be in an empty partition, the result is the first record in the next non-empty partition, e.g., the result of query=15 is record 3. To make the Shift-Table layer consistent for the empty partitions, we put pseudo values for $\Delta,C$ in the mapping layer such that they refer to the same range as the next existing partition. If $P_{k^\varnothing}$ is an empty partition and $P_{k}$ is the first non-empty partition after $P_{k^\varnothing}$, then $C_{k^\varnothing} = C_k$ and $\Delta_{k^\varnothing} = \Delta_k + (k - k^\varnothing)$. The pseudo $\Delta,C$-values are depicted using dashed arrows in Figure~\ref{fig:direct-mapping}.

\subsection{CDF and duplicate values}
\label{subsec:duplicates}
It should be noted that the \textit{empirical CDF} function, i.e., $F(X) = P(X\leq x)$ does not exactly identify the result of a range query on x. In this paper, we use the CDF (F(x)) notation as the index of the result corresponding to $x$. We consider range queries of type ($\texttt{key} <= \texttt{query}$), hence the CDF for a point $x$ is the relative position of the \textit{first} key in the indexed keys, as the range is scanned towards the right. More precisely, we assume that $NF(x_0) = 0$  and $NF(x_{N-1}) = N-1$ (for the last key). 

A range learned index built for a specific comparison operator, say $x\leq q$, can be used for other operators ($\geq, >,$, etc.) with a brief left/right scan. However, if there are too many duplicates in the indexed data, then the the performance of the learned index will be worse for queries that do not match the presumed definition of F(X). In such cases, it is more efficient to use the specific definition of $F(x)$ that reflects the position of the result of the query in the most common type of constraint in the queries. For example, if most of the queries are of type $x>=q$, then $F(x)$ should be defined such that $NF(x)$ identifies the index of the last key among the duplicate values.

\subsection{Building the Shift-Table layer}
Algorithm~\ref{alg:dm-build} describes how the mapping of the Shift-Table layer is built. In the first stage, it computes 
the $\Delta,C$ values and updates for the non-empty partitions, i.e., $P_k$s for which at least one of the indexed keys is mapped to $k$. In the second stage, a backward traversal is performed on the Shift-Table layer and the compute the pseudo-values for the empty partitions (Algorithm~\ref{alg:dm-build}, lines 10--14). Starting from the last entry, a pseudo-partition has the same count ($C$) as the first non-empty partition on its right side, but the shift $\Delta$ is adjusted so that they both point the the same region for local search.

The computational complexity of building the Shift-Table layer is $O(N) \times O(F_\theta)$ to compute the drifts and updating the mapping, as it only traverses the data and the Shift-Table layer once. In case that running the model is expensive, model executions can be parallelized for faster execution.

\begin{algorithm}
    \caption{Building the Shift-Table layer}
    \label{alg:dm-build}
    \begin{algorithmic}[1] %
        \Procedure{Shift-Table\_Build}{model ($F_\theta$), data} 
        \State Shift-Table = Array of tuples \texttt{<}$\Delta,C$\texttt{>}, all set to zero
        \ForAll{x $\in$ data} 
            \State $pos = NF(x)$ \Comment{Position of x (sec~\ref{subsec:duplicates})}
            \State  $k = [NF_\theta(\text{data}[\text{i}])]$
            \State $\Delta = pos - k$
            \State Shift\_Table[k].$\Delta$ = min$\left( \text{Shift\_Table}[k].\Delta , \Delta \right)$
            \State Shift-Table[k].C += 1
        \EndFor

        \For{$k \gets N-1 \cdots 0$}
            \If{Shift\_Table[k].C = 0} \Comment{Empty partitions}
                \State Shift\_Table[k].C = Shift\_Table[k-1].C
                \State Shift\_Table[k].$\Delta$ = Shift\_Table[k-1].$\Delta + 1$ 
            \EndIf
        \EndFor

        \State \Return Shift\_Table
        \EndProcedure
    \end{algorithmic}
\end{algorithm}

\subsection{Compressing the Shift-Table layer}
\label{subsec:Shift-Table-compression}
Correcting the prediction of the model using the Shift-Table layer takes a single DRAM lookup irrespective of the size of the index. However, it might be of interest to reduce the size of the layer. The Shift-Table layer is an array of size N, containing $\texttt{<}\Delta,C\texttt{>}$ tuples. Further compression can be used to decrease the memory footprint of the Shift-Table layer. 

One approach is to keep a single parameter instead of the $\texttt{<}\Delta,C\texttt{>}$ tuples. In this regard, a predicted position $k$ should be mapped to the key that is in the median point among the keys in $P_k$, which is 

\begin{equation}
    \Bar{\Delta}_k = \left[\Delta_k + \frac{C_k}{2}\right]
\end{equation}

To correct using the $\Bar{\Delta}_k$ values, the final position is computed as  $pos = k + \Bar{\Delta}_k $, which indicates where the search should be started without specifying the guaranteed range that should be searched. Therefore, search algorithms that require the boundaries specified such as binary search cannot be used for local search. As discussed in section~\ref{subsec:motivation-difficulty}, linear or exponential search can be used for local search without boundaries, but they are slightly slower if the error is considerable after the correction.

A second approach that complements the first one, is to shrink the size of the Shift-Table layer by merging nearby partitions. We can extend the definition of $\mathcal{P} = \left\{P_1,\cdots,P_N\right\}$ to allow partitions that have a size of $M < N$. We define $M$ partitions $\mathcal{P}^M = \left\{P^M_1,\cdots,P^M_{M}\right\}$ where each partition is defined as:
\begin{equation}
     P^M_k = \{x \; | \; [M F_\theta(x)] = k\} 
\end{equation}

Similarly, $\Delta^M_k$ is the minimum "move to the right" shifts that each of the keys in $P^M_K$ need:

\begin{equation}
    \Delta^M_k = \text{min}\left(N F(x) - [NF_\theta(x)]\right)\;\; \forall x \in P^M_k
\end{equation}

and $C_k$ should be defined such that the boundary is valid for all keys in $P^M_K$, which is:

\begin{equation}
    C^M_k = \text{max}(N F(x) - (%
    \underbrace{[NF_\theta(x)] + \Delta^M_k}_{\text{\tiny start of the search window}}%
    ))\;\; \forall x \in P^M_k
\end{equation}

To combine approaches to compact the Shift-Table layer, we can use average drifts $\Bar{\Delta}^M_k$ instead of the $\texttt{<}\Delta^M_k,C^M_k\texttt{>}$ pairs:
\begin{equation}
    \Bar{\Delta}^M_k = \left[\frac{1}{\vert P^M_k \vert}\sum_{x \in P^M_k}\left(N F(x) - [NF_\theta(x)]\right)\right]
\end{equation}

and then use $[NF_\theta(x)] + \Bar{\Delta}^M_{[MF_\theta(x)]}$ as the corrected prediction. Suppose the same data as in Figure~\ref{fig:direct-mapping}, but instead of a Shift-Table layer of size N, we use only M=30 partitions.  Table~\ref{tab:dm-correction-m30} shows how a compact Shift-Table layer is built and used for correction, on a portion of the index. We use the same model ($F_\theta=[x / 1000]$), hence the prediction is $NF_\theta(x)=[0.1 x]$, and the partition corresponding to a key is $NF_\theta(x)=[0.03 x]$. All of the records from data[35..39] are assigned to the same partition $P_{23}^{30}$ and their predictions are shifted 40 records backwards. Note that when $M\neq N$, a partition does not specify a single point (or range) for all of the keys in the partition. 
Instead, the position of a key after correction depends on both $NF_\theta(x)$ (prediction) and $MF_\theta(x)$ (partition number). For example, all keys belonging to $P^{30}_{23}$, i.e., data$[35 \cdots 39]$ have the same correction of $\Bar{\Delta}^{30}_{23} = -40$, but their final predictions are different. Therefore, the correction error of a compact Shift-Table layer is less than the number of elements in the partitions. 

\begin{table}[htbp]
  \centering
  \caption{Illustration of Shift-Table with $M=30$ mapping entries on an index with $N=100$ keys}
  \resizebox{\columnwidth}{!}{%
    \begin{tabular}{l|r|r|r|r|r|r|r|r}
        Index &  34    & 35    & 36    & 37    & 38    & 39    & 40    & 41 \\ \hline
        key (x)  &  752   & 769   & 770   & 771   & 782   & 785   & 820   & 830 \\ \hline
        Predicted index= [0.1 x] &  75    & 76    & 77    & 77    & 78    & 78    & 82    & 83 \\ \hline
        Error before correction    & -41   & -41   & -41   & -40   & -40   & -39   & -42   & -42 \\ \hline
        Partition (k) = [0.03 x] &  22 & \multicolumn{5}{c|}{23} & \multicolumn{2}{c}{24} \\ \hline
        $\Bar{\Delta}_k^{30}$    & -41   & \multicolumn{5}{c|}{-40}   & \multicolumn{2}{c}{-42} \\ \hline
        Prediction after correction & 34    & 36    & 37    & 37    & 38    & 38    & 40    & 41 \\ \hline
        Error after correction     & 0     & 1     & 1     & 0     & 0     & -1     & 0     & 0 \\
        \end{tabular}%
    }
  \label{tab:dm-correction-m30}%
\end{table}%

The drift of $P^M_k$, namely $\Bar{\Delta}^M_k$ is the index of the median key among the members of $P^M_k$. This means that if the key is predicted to be in the $k$'th partition (among the $M$ partitions), the local search is done around $[NF_\theta(x)] + \Bar{\Delta}^M_k$. 

Using a Shift-Table layer of size $K < N $ does not affect the complexity of building the layer, which is $O(N) \times O(F_\theta) + O(M)$. However, if the midpoint-values are used (correction without specifying the boundary), it is possible to construct the map using a sample of the indexed keys, which comes at the cost of the accuracy. Using a sample of size $S < N$, the layer can be built in $O(S) \times O(F_\theta) + O(K)$ time. 

Nonetheless, keep in mind that the Shift-Table layer is designed for applications that favour latency to memory footprint, hence reducing the memory footprint of the Shift-Table layer by a large factor will limit its margin for improvement as the fine-grained details of the empirical CDF will be lost to some extent.

\subsection{Measuring the error}
\label{subsec:measuring-dm-error}
 Since the Shift-Table layer specifies a range for local search, the notion of error is not trivial. However, we can use the estimates without range $\Bar{\Delta}$), for which the correction picks the median value among the keys in the $P_k$. The error for the keys in each partition is $\left\{[\frac{C_k}{2}],\cdots,0,\cdots,[\frac{C_k}{2}]\right\}$ if $C_k$ is odd, and $\left\{[\frac{C_k}{2}]-1,\cdots,0,\cdots,[\frac{C_k}{2}]\right\}$ if $C_k$ is even. The average error is approximately $C_k/4$. %

In a learned index without Shift-Table, the error is the distance between $F(x)$ and $F_\theta(x)$. After correcting the model with the Shift-Table, however, the error only depends on the $C_k$ values, i.e., a prediction error only occurs when $[F_\theta(x)]$ predicts the same position for multiple keys. Therefore, the local search range and the error are combinations of multiple step functions over the $P_k$s with $C_k > 1$.  

The average error depends on the data distribution in the query workload. If the queries are uniformly sampled from the keys, then the average error is:
\begin{equation}
    \label{eq:error-of-dm}
    \Bar{e} = \frac{1}{2N} \sum_{k\in \mathcal{P}}{C_k^2}
\end{equation}

\subsection{Behaviour of the Shift-Table layer}

\begin{figure}
\centering
  \begin{minipage}[b]{.49\linewidth}
    \centering
    \includegraphics[width=.85\linewidth]{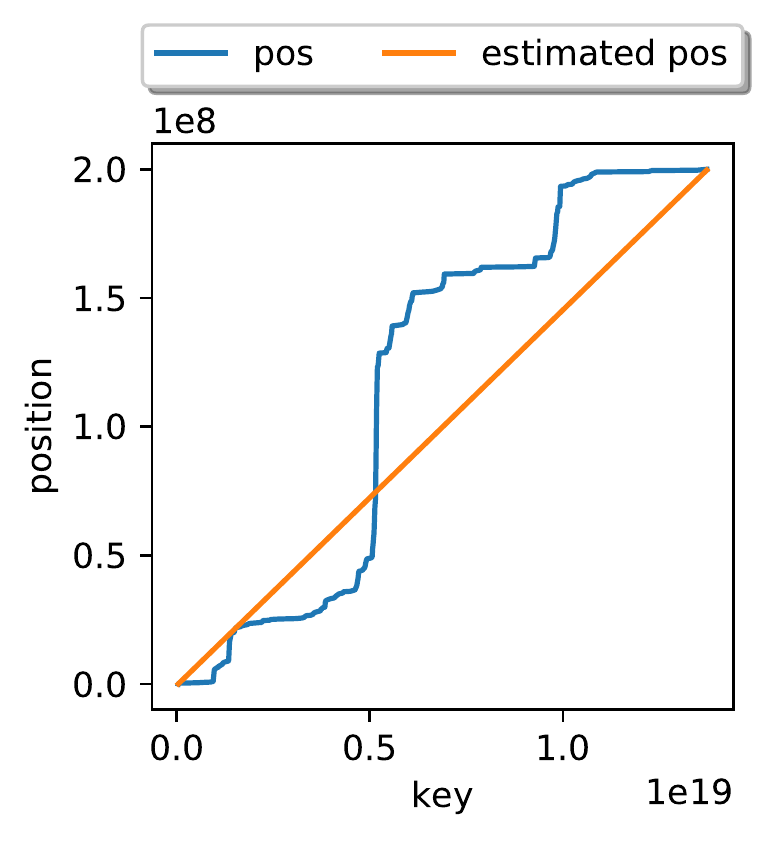}
    \subcaption{Example data \& model}
    \label{fig:direct-mapping-example-CDF}
  \end{minipage}
  \hfill
  \begin{minipage}[b]{.49\linewidth}
    \centering
        \includegraphics[width=.95\linewidth]{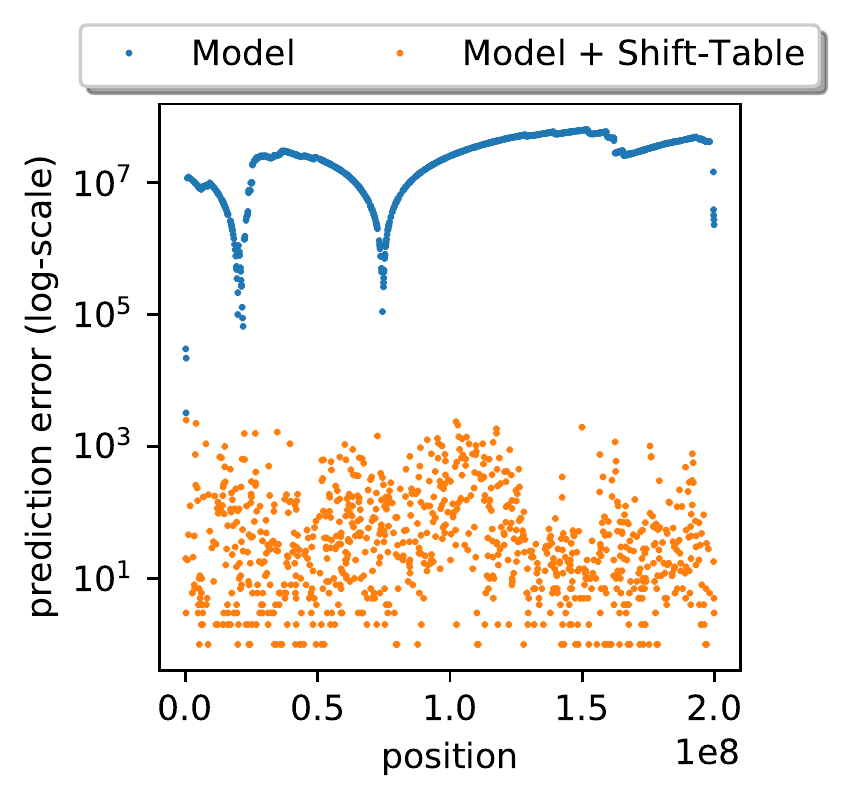}
    \subcaption{Error}
    \label{fig:direct-mapping-example-correction}
  \end{minipage}
    \caption{Error correction using the Shift-Table layer}
    \label{fig:direct-mapping-example}
    \vspace{-10pt}
\end{figure}

Figure~\ref{fig:direct-mapping-example} illustrates how the Shift-Table layer corrects the error of a linear interpolation model on the OSMC data. While the model is too simple to capture the patterns in data, the Shift-Table layer alone is effective for correcting the predictions. While the average error of the model is 28 million keys, Shift-Table reduces the error to only 129 keys.

\missingfigure{Add more subfigures to Fig.~\ref{fig:direct-mapping-example} to show that a learned model could be more effective in reducing error}

Shift-Table corrects two types of error. First, when the model has a considerable local bias, which means that $NF(x)$ diverges significantly from $NF_\theta(x)$ in a sub-range of the data distribution. 
The second type of error is the fluctuations of the distribution between the nearby keys, for most of which the Shift-Table layer is very effective. The only type of error that can degrade the performance of the Shift-Table layer is when there is a congestion of keys in a small sub-range of values, leading to many of the keys being classified in a single layer, and hence having some partitions with high $C_k$.

The behavior of the Shift-Table layer and its error estimate indicates that it can be effective in eliminating different types of errors that models have. One common type of error is the local bias in the model, i.e., when the error of the model, i.e., $NF_\theta(x) - NF(x)$ has a considerable \textit{bias} in some sub-ranges of the distribution, meaning that the $F$ and $F_\theta$ diverge at some point. This happens when the model cannot capture the CDF in a local neighborhood. Table~\ref{tab:sosd_results} shows that even if a single line is used as a model, which has a huge bias in most areas of the distribution, the Shift-Table layer can efficiently eliminate the huge bias of a fully linear model (a single line as a model), and reduces the error significantly such that the linear model outperforms all other algorithms for the real-world datasets, as well as the uspr dataset (sparse uniformly-distributed integers) which has a significantly higher variance than uniformly-distributed dense integers.

Another type of error that the Shift-Table layer eliminates is the local variance in the data, which is the fluctuations of the values between nearby keys. This type of error is very common in real-world data. For example, the \textit{face, uspr, and uden} datasets all follow a uniform distribution, but they have different local variances, which is the amount of fluctuations in the nearby keys. The uden dataset is very easy to model using the learned indexes and does not require a helping layer such as Shift-Table. The other two datasets, however, are very hard to model using the learned index structures. This is because the Shift-Table model can easily correct the fluctuations of values (different increments between each two points), as long as the model does not predict a single record for a lot of nearby keys (resulting in a high $C_k$ value).

\subsection{Cost model of the Shift-Table layer}
\label{subsec:cost-of-dm}
The accuracy of the model after correction with Shift-Table depends on the cardinalities of the partitions ($C_i$ values). Ideally, if the records of each partition reside on a single cache line, the results will be retrieved in a single memory lookup. The cost of local search, i.e., the mapping between the accuracy in each partition and the latency to do local search depends on the hardware. As discussed in section~\ref{subsec:cost-of-local-search}, the latency of search for various ranges can be measured by a micro-benchmark over non-cached regions with different sizes. Let $L(s)$ be the measured latency of non-cached search over a range containing $s$ records. The latency for looking up a key in a region of size $s$ is $L(C_k)$. Assuming that the queries have the same distribution as the data points, the average lookup latency for the index is:

\begin{equation}
    \label{eq:latency-dm}
     \text{Latency with Shift-Table = Latency}(F_\theta) + \frac{1}{N} \sum_{k\in \mathcal{P}}{C_k L(C_k)}
\end{equation}

The cost model can also be used to estimate which of the local search algorithms should be used, by substituting in equation~\ref{eq:latency-dm} the local search cost of each local search algorithm, i.e., $L(s)$ mappings for linear, binary, and exponential search; and for and their different implementations. Branch-optimized binary search would be the natural choice if the Shift-Table model can determine the boundary (if using the $\Delta_k,C_k$ pairs), otherwise either linear or exponential search should be chosen based on the latency estimate.  

Taking the cost of running the Shift-Table layer into account, we should consider how much the correction improves the accuracy of the learned index model and hence estimate the speedup. The lookup time of the model without using the Shift-Table model can be estimated once the Shift-Table model is built, without running a speedup benchmark. The model error for each key is $\Bar{\Delta}_k = \Delta_k + \frac{C_k}{2}$, therefore the estimated runtime of the index without correction is:

\begin{equation}
    \label{eq:latency-dm-raw}
     \text{Latency without Shift-Table = Latency}(F_\theta) + \frac{1}{N} \sum_{k\in \mathcal{P}}{C_k L(\Bar{\Delta}_k)}
\end{equation}

\subsection{CDF model validity constraint}
\label{subsec:cdf_model_validity}
The correction layer requires the learned model to be a valid CDF function, i.e., $F_\theta(x)$ should be monotonically increasing: $x_i > x_j \longrightarrow F_\theta(x_i) >= F_\theta(x_j)$. Among our baselines, the RadixSplines learned index always produces a valid (increasing) CDF, but the RMI index does not always produce monotonically increasing predictions. In RMI, for example, the CDF model might decrease when using cubic models~\cite{marcus2020cdfshop} or on the edge point between two models in the second-level. If $F_\theta(x)$ is not monotonically increasing, then the correction layer could identify a range that does not include the query result, because the values of $x$ for which the learned model predicts $k$'th record are not in a contagious memory block. 

A learned index model that is non-monotonic can still use the Shift-Table layer, as the output of the Shift-Table layer would still predict a position but it is not guaranteed that the position is in the predicted range. Therefore, the local search algorithm should check if the query is in the predicted range and perform a search outside of the range. Another hack for non-monotonic model is to use the $\Bar{\Delta}$  midpoint-values instead of the $\Delta_k,C_k$ pairs, which predicts a location (instead of a range) to start the local search. %

If the Shift-Table layer uses the $\texttt{<}\Delta^M_k,C^M_k\texttt{>}$ pairs, it can determine the range for local search and we can apply either linear or binary search, depending on the error range. We do linear search if the range is smaller than a threshold (8 keys, in our experiments), otherwise a binary search is performed. However, if it only contains the average shift values ($\Bar{\Delta}_k$, it predicts a position without specifying the boundaries that contain the record; hence either linear or exponential search can be performed depending on the \textit{average} error rate and performance objectives (average or worst-case latency).

\subsection{Tuning the system}
\label{subsec:tuning}
\todo[inline]{This section was written when we had two layers. Now that we have only one layer (Shift-Table), we might need to make it brief and merge it with another section}
The Shift-Table layer is optional and adds overhead to the search. Therefore, enabling Shift-Table is only worthwhile if it can eventually accelerate the original learned index structure. An effective configuration of the index is a choice between  1) Using the model alone, 2) model + Shift-Table. Note that the Shift-Table layer is optional and can be deactivated with zero cost. The output of the model and the Shift-Table layer are of the same type and both represent a prediction of the records, hence if the Shift-Table layer is disabled, we can easily use the model alone for prediction of the records.

While tuning the system, the performance of each configuration can be directly measured using performance tests, or by measuring the model error and then using the cost model of the Shift-Table model on the bottom of the architecture (section~\ref{subsec:cost-of-dm}).

The parameters of the architecture, i.e., the Shift-Table array size $M$ and the parameters of the learned CDF model, can be tuned by computing the error estimate using Shift-Table's cost model, or alternatively, by running a performance tests on the built architecture. Our suggested default value for the Shift-Table layer is $M=N$, because using a mapping layer that has the same number of entries as the keys will ensures that the layer can exhibit its ultimate effect to eliminate the signed error, and does not have more latency compared to using smaller $M$ values. 

An advantage of Shift-Table is that the learned model does not need to be very accurate, as a correction will be applied anyway. Therefore, a more relaxed measure can be used instead of least-square error. In this paper, however, we do not learn the model w.r.t. the Shift-Table layer, for the sake of simplicity and to keep the Shift-Table layer detacheable (optional), preserving the assumption that the Shift-Table layer can be disabled to free up memory space on run-time while the model can still be used.

The accuracy of the learned model also determines the size of the entries of the Shift-Table layer. Each mapping entry should at most fit a $\Delta$ value of $\Delta_{MAX}$, which is the maximum error of the model. If, for example, the error is smaller than $2^16/2$, then a 16-bit integer (\texttt{short} type) can be used.

\section{Evaluation}
\label{sec:evaluation}
In this section, we compare the performance of our proposed method with the SOSD benchmark~\footnote{\url{https://github.com/learnedsystems/SOSD/tree/mlforsys19}}, which is a recent benchmark for search on sorted data. The benchmark includes learned indexes, classical indexes, and no-index search algorithms.  

\iparagraph{Experimental Setup}
The algorithms are implemented in C++ and compiled with GCC 9.1. The experiments are performed on a system with 16 GB of memory and Intel Core i7-6700 (Skylake), which has four cores and is running at 3.4 GHz with 32 KB L1, 256 KB L2, and 8 MB L3 caches. The operating system is Ubuntu 18.04 with kernel version 4.15.0-65. In our setup, the LLC miss penalty measured by Intel Memory Latency Checker~\footnote{\url{https://software.intel.com/en-us/articles/intelr-memory-latency-checker}} is 36 ns, which is the minimum lookup time of an ideal index. 

Note that all data resides in main memory. The range index finds the first indexed key that is equal to or bigger than the lookup key. Also, the keys on the physical layout are sorted (i.e., it is a clustered index), so that the entire result set of the range query can be returned once the first key is found. Similar to~\cite{kraska2018case,kipf2019sosd}, we only report the lookup time for the first result and do not include the scan times in our experiments because all indexes use the same layout for the data records.

\iparagraph{Datasets}
For the sake of reproducibility, we used the same datasets as in the SOSD benchmark, which contains four datasets synthetically generated from known distributions and four real-world ones.
The synthetic data are generated from different distributions, namely 
\textit{logn}: lognormal distribution $(0,2)$, 
\textit{norm}: normal distribution, 
\textit{uden}: uniformly-generated dense integers, 
and \textit{uspr}: uniformly-generated sparse integers. 
The real-world datasets are \textit{face}: Facebook user IDs~\cite{van2019efficiently},
\textit{amzn}: book sale popularity from Amazon sales rank data%
~\footnote{\url{https://www.kaggle.com/ucffool/amazon-sales-rank-data-for-print-and-kindle-books}}, 
\textit{osmc}: uniform sample of OpenStreetMap locations%
~\footnote{\url{https://aws.amazon.com/public-datasets/osm}}, 
and \textit{wiki}: timestamps of edit actions on Wikipedia articles%
\footnote{\url{https://dumps.wikimedia.org}}. All datasets contain 200M unsigned integers.

\begin{table*}
\centering
\caption{Comparison of lookup times (nanoseconds per lookup) with the SOSD benchmark. The red box indicates the base model (IM) and the enhanced versions.}
\vspace{-10pt}
\includegraphics[width=0.99\linewidth]{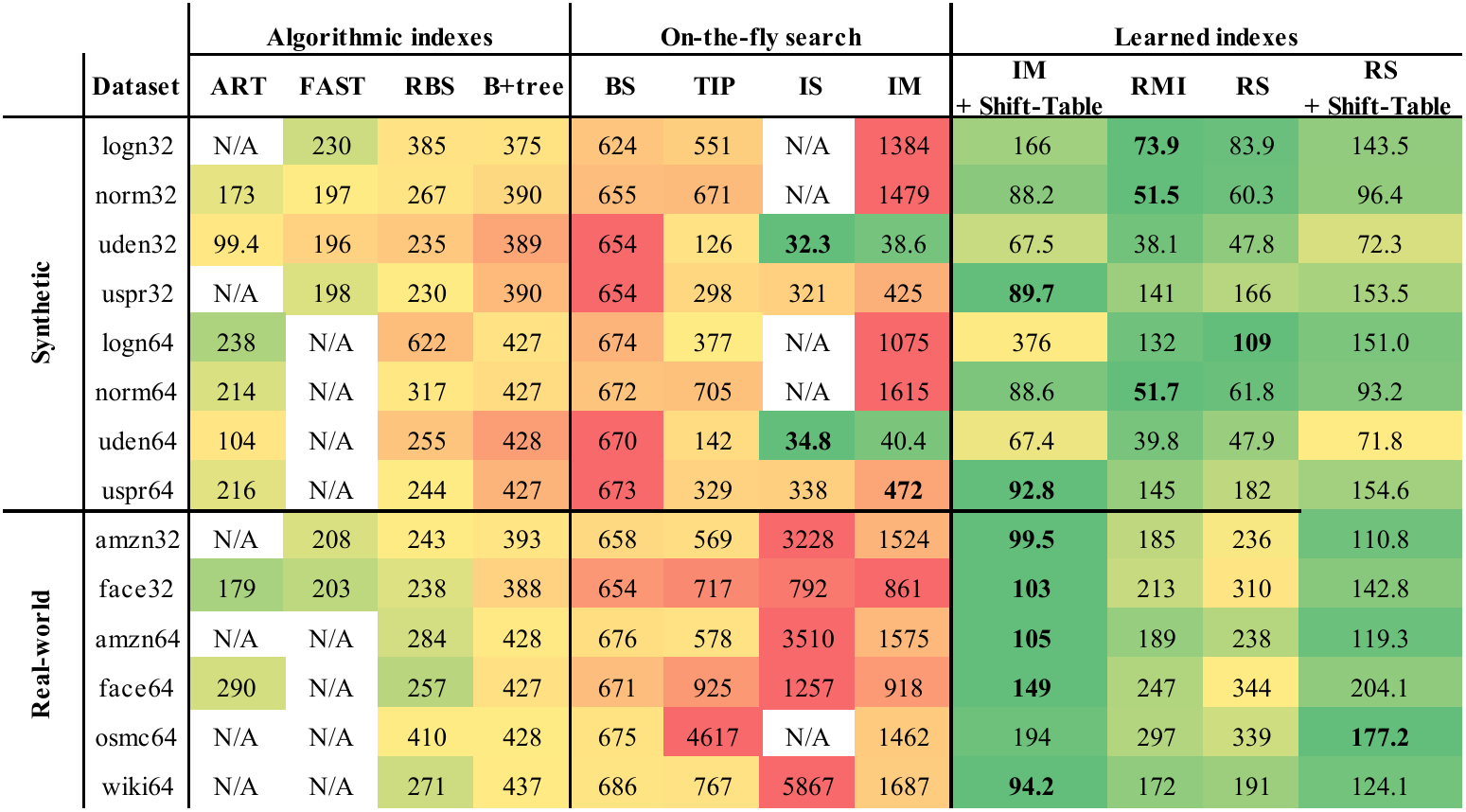}
\label{tab:sosd_results} 
\vspace{10pt}
\end{table*}

\iparagraph{Implementation details}
Our experiments are based on the SOSD benchmark~\cite{kipf2019sosd}. The baseline includes two learned indexes, namely RadixSpline~\cite{neumann2008smooth} (RS), which uses linear splines; and Recursive Model Index (RMI), which uses a hierarchy of models. Note that  RMI has a choice of different models and SOSD~\cite{kipf2019sosd} specifically handpicked the best models for each of the datasets in the benchmark~\footnote{The architectures and parameters of the RMI models used for each dataset is specified at \url{https://github.com/learnedsystems/SOSD/blob/mlforsys19/scripts/build_rmis.sh}}. SOSD also includes no-index search algorithms such as binary search (BS), linear interpolation search (IS), and the recently suggested non-linear triple-point interpolation (TIP)~\cite{van2019efficiently}. We also compare against algorithmic index structures such as ART: Adaptive Radix Tree~\cite{leis2013adaptive}, FAST~\cite{kim2010fast}, RBS (Radix Binary Search): a two-stage algorithm in which a radix structure that maps a fixed-length key prefix to the range of all keys having that prefix and then a binary search is performed on the range~\cite{kipf2019sosd}, and STX implementation of B+tree~\cite{stx2007}. Finally, we included four \textit{On-the-fly} search algorithms, namely BS: Binary search (STL implementation), TIP: three-point interpolation~\cite{van2019efficiently}, Interpolation search, which is similar to binary search but uses interpolated positions in each iteration, and IM: \textit{Interpolation as a Model}: a dummy model that interpolates the key between the minimum and maximum value of the keys and then performs exponential search around the predicted key.

The experiments use either 32- or 64-bit unsigned integer IDs for the key (depending on the dataset), and 64-bytes for the payload.

\subsection{The SOSD benchmark}
\label{subsec:eval_sosd_benchmark}
To test the effectiveness of the suggested layers compared to learned indexes, we use a simple interpolation model (IM), i.e., $F_{\theta}(x) = (x - minVal)/(maxVal-minVal)$. Such a dummy model is deliberately chosen to purely delegate the burden of data modelling to the correction layers.

\todo[inline]{Todo: This paragraph was written for 2 layers in mind. Updated, but needs to be revised.}
The Shift-Table layer has the same number of entries as the actual data, i.e., $M=N$. We followed the tuning procedure discussed in section~\ref{subsec:tuning}: we start from the model (IM and RS) and consequently evaluate IM+Shift-Table and RS+Shift-Table. The cost of running the Shift-Table layer is around 40ns, which pays off by reducing the prediction error and thus lookup time. Therefore, based on the cost model of the Shift-Table layer (Section~\ref{subsec:cost-of-dm}) and the error-to-latency micro-benchmark (Figure~\ref{fig:binary_search_vs_local_search_time}), we should not add the Shift-Table layer if the error before adding the configuration is less than a threshold (10 records), or 2) the error of the index after adding the Shift-Table layer does not decrease by a factor of 10 (roughly equivalent to the 50-nanoseconds latency the additional layer, according to the error-to-latency micro-benchmark).

Table~\ref{tab:sosd_results} compares the lookup times (nanoseconds per lookup) of the baseline algorithms with our dummy interpolation model (IM), and the two corrected versions, i.e.,  IM+Shift-Table and RS+Shift-Table. Note that ART does not support data with duplicate keys, and FAST does not support 64-bit keys. Also, interpolation search (IS) takes too much time on some datasets, because the execution time of interpolation search highly depends on the uniformity of data distribution, varying from O(loglogN) + O(1) iterations on uniform distributions, to O(N) iterations for very skew ones~\cite{van2019efficiently}.

For the synthetic datasets, the difficulty of the datasets for our dummy linear interpolation model varies from very easy (uden64) to extremely hard (logn64). While the Shift-Table layer significantly improves a dummy layer on non-uniform data distributions, it cannot outperform the learned index models. This is not surprising, as all synthetic datasets (uniform, lognormal, and uniform) have a pattern derived from continuously differentiable density functions, hence the distribution is similar to a straight line on smaller sub-ranges as we "zoom in" the data distribution (e.g., see Figure~\ref{fig:lognormal}). Therefore, a learned index structure composed of linears at the bottom (including both RMI and RS) can effectively model the distribution using a very compact representation. 

For the real-world data, however, the fluctuations in data severely affect both RMI and RS learned indexes. The Shift-Table layer, effectively corrects a highly inaccurate dummy IM model, such that it outperforms the RMI learned index by 1.5X to 2X on all datasets, while RS falls behind both. Keep in mind that RMI requires to be tuned with the best architecture and parameters, while Shift-Table does not require a manual training process and can even work with a simple model such as IM that is not trained, and yet deliver a lower latency.

\begin{figure}
    \centering
    \includegraphics[width=.9\columnwidth]{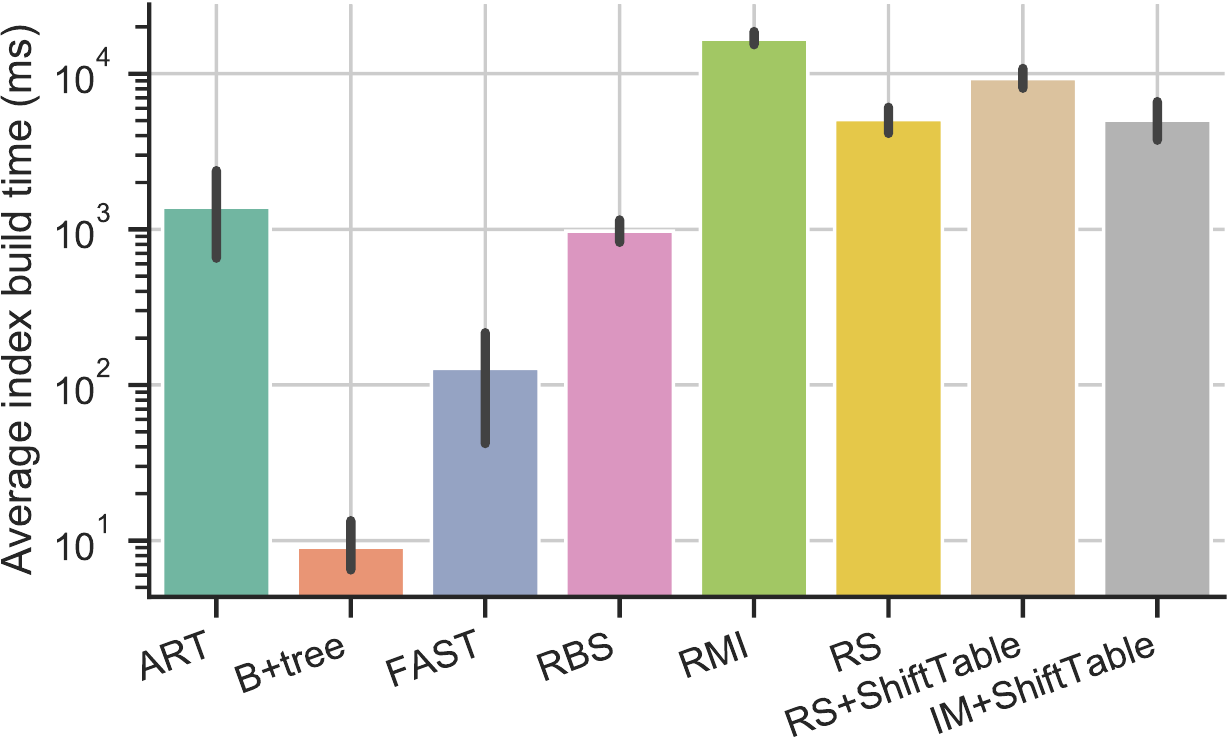}
    \caption{Build times (average time for all datasets)}
    \label{fig:build_times}
    \vspace{-10pt}
  \end{figure}

Figure~\ref{fig:build_times} shows the average build times of the indexes, along with the standard deviation bars indicating how the build time varies for different distributions. Please note that the RMI implementation used in the SOSD benchmark needs to be compiled for faster retrievals, however we did not include RMI's extra overhead for compiling the code and only reported the build time. IM+Shift-Table, the winner method latency-wise, also takes either the same or even less build time than the competing learned indexes.

\subsection{Explaining the performance}
\label{subsec:eval-performance-details}
The latencies reported in Table~\ref{tab:sosd_results} present the fastest configuration for each learned index. In this section, we present the details of the tuning process to see the optimum performance of each learned index.

For those indexes that have a parameter affecting the index size (such as the branching factor in B+tree, and the number of radix bits in ART, RS, and RBS), the performance can be tuned by evaluating the latency for different index sizes. 

\begin{figure*}
    \centering
    \includegraphics[width=\linewidth]{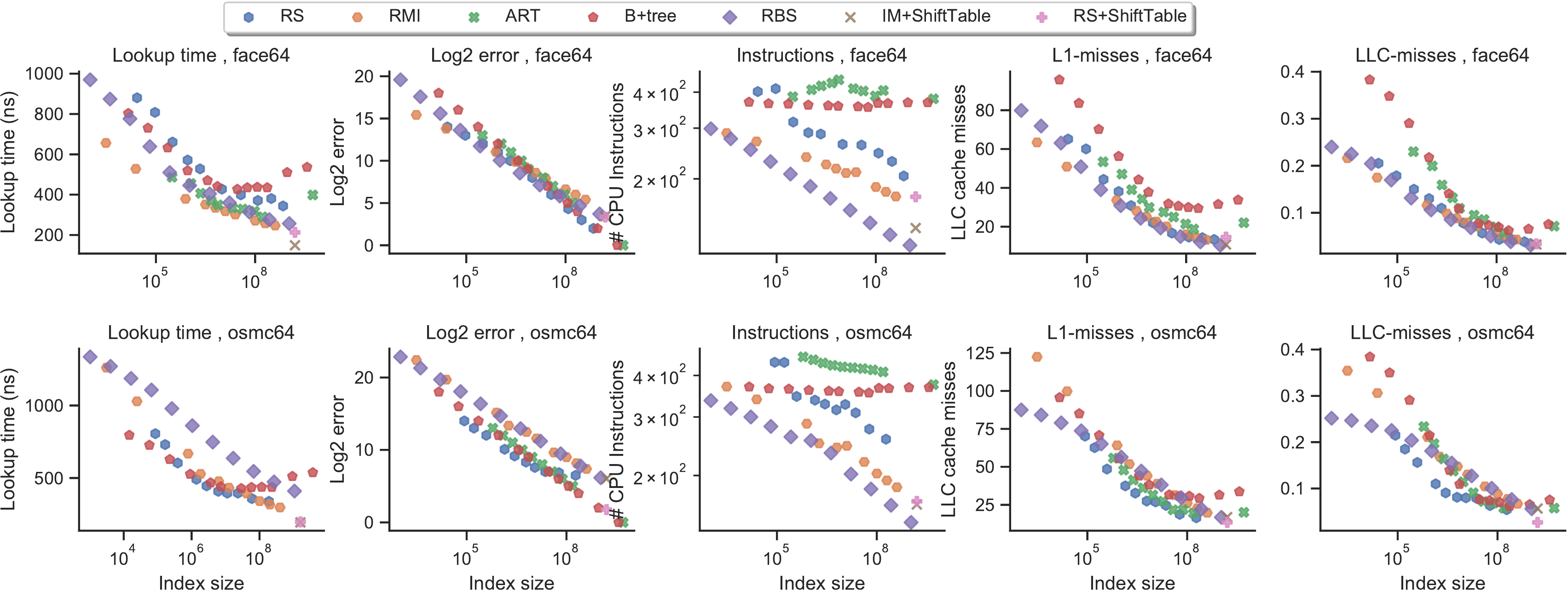}
    \caption{Analysis of the effect of index size on performance}
    \label{fig:various_index_sizes}
  \end{figure*}

Figure~\ref{fig:various_index_sizes} shows the latencies of the indexes for the face64 and osmc64 datasets, along with the average Log2 error,  CPU instructions, and L1/LLC cache misses. 
IM+Shift-Table and RS+Shift-Table achieve faster lookup times on both datasets. For most indexes, except RMI and RBS, the latency does not improve beyond a certain optimum index size, after which the latency increases again. RBS has a much larger latency than both [IM/RS]-Shift-Table indexes of the same size, and extrapolating the RMI latencies also suggest that if we could extend RMI size to 1400MB (equal to Shift-Table's size), it could not achieve a game-changing performance on either of the datasets. Note that we could not run RMI with larger models because RMI embeds the parameters into the code, and the compile times for models larger than 400MB were astonishingly high. 

Average Log2 errors indicate the average number of iterations in binary search for the last-mile search stage. Larger models result in lower Log2 errors in all indexes and lead to faster last-mile search, however, once the model exceeds the LLC cache sizes, cache-miss rate increases (when running the model), and hence the prediction time worsens. For RS, ART, and B+tree, the cache misses and extra overhead of running the models increases either the number of instruction, the cache misses, or both, enough to prevent the index from improving latency by increasing the footprint.

\subsection{Layer size}

\begin{figure}
\centering
  \begin{minipage}[b]{.48\textwidth}
    \centering
    \includegraphics[width=\linewidth]{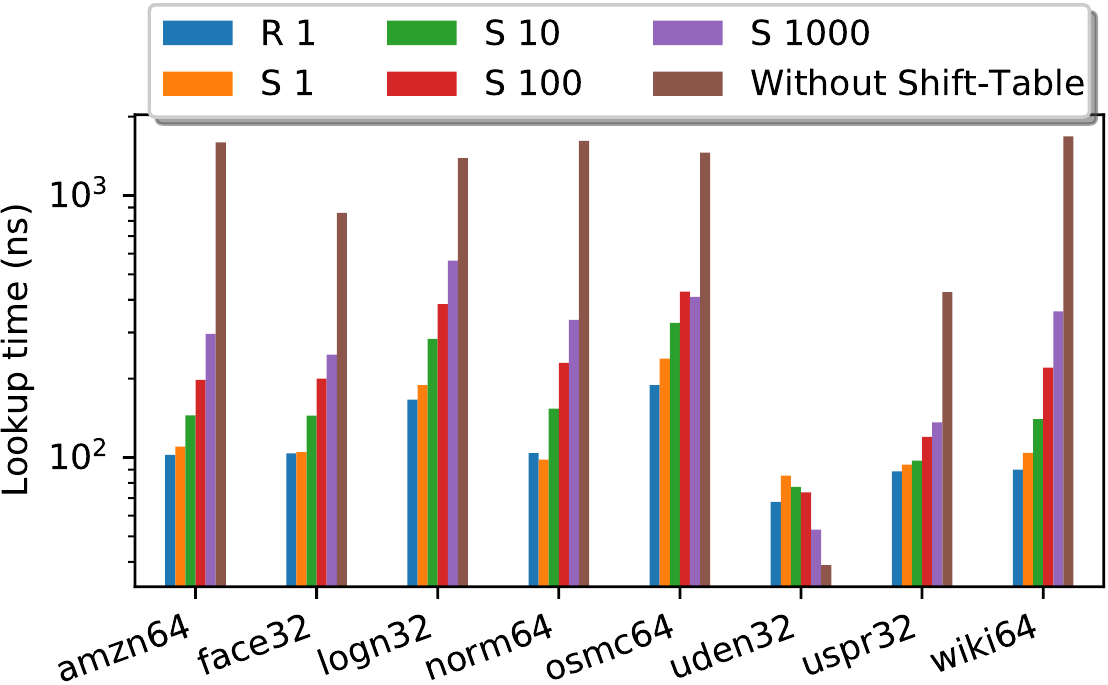}
    \subcaption{Latency}
    \label{fig:Shift-Table_analysis_latency}
  \end{minipage}
  \hfill
  \begin{minipage}[b]{.48\textwidth}
    \centering
        \includegraphics[width=\linewidth]{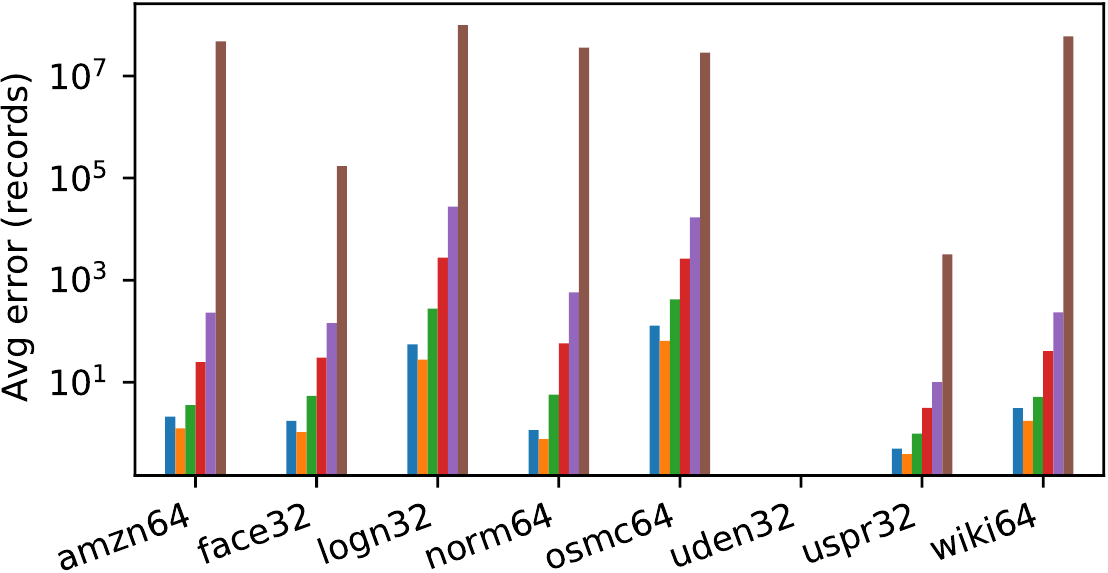}

    \subcaption{Error}
    \label{fig:Shift-Table_analysis_error}
  \end{minipage}
    \caption{Analysis of the effect of Shift-Table layer size}
    \label{fig:Shift-Table_analysis}
    \vspace{-20pt}
\end{figure}

As discussed in section~\ref{subsec:Shift-Table-compression}, the Shift-Table layer can be compressed by merging multiple entries, hence reducing its footprint. Figure~\ref{fig:Shift-Table_analysis} shows the effect of the Shift-Table layer size on lookup time and prediction error. Shift-Table can operate in two modes: \textbf{R-1}: a full layer containing $\texttt{<}\Delta^M_k,C^M_k\texttt{>}$ pairs similar to Figure~\ref{fig:direct-mapping} that indicates the exact \textit{range} for local search (hence enabling binary search); and \textbf{S-X}: a compressed single-entry map similar to Table~\ref{tab:dm-correction-m30} containing one  $\Bar{\Delta}^M_k$ entry per $X$ records. Thus, S-X contains $M=N/X$ entries; and the memory footprint of S-1 is half the size of R-1. 

The error of the S-1 Shift-Table is slightly more than that of R-1. This is due to the fact that S-1 is designed to draw boundaries for binary search; hence it always points to the first record of each partition; while R-1 always points to the middle of the partition and almost half the error of S-1. Performance-wise, however, S-1 always has the lowest latency, because its boundaries for the last-mile search operation do not need to be discovered using additional boundary-detection algorithms such as exponential search. As expected, compressing the Shift-Table by allocating one entry per $X$ records increases the error and hence degrades the performance. This is due to the fact that with higher compression ratios, the ability of Shift-Table to "memorize" the fine-grained details of the data distribution degrades due to the loss of information after merging.

\section{Related work}
\label{sec:relatedwork}

\textbf{On-the-fly search on sorted data} 
A fundamental problem that is studied for decades is how to find a key among a sorted list of items. The classic approach is binary search and numerous extensions have been suggested to improve it for special cases, most notably interpolation search~\cite{peterson1957addressing} and exponential search~\cite{bentley1976almost}. For data distributions that are close to uniform, interpolation-search is shown to be very effective~\cite{price1971table,graefe2006b,van2019efficiently}. Due to the growing gap between CPU power and memory latency in the past decade, more advanced interpolation techniques such as three-point interpolation are becoming viable on modern hardware~\cite{van2019efficiently}. Exponential search enables binary search over an unbounded list. Exponential search is also extensively used in learned indexes when the key is more likely to be near a "guessed" location, but a guaranteed boundary around the guessed point that contains the data is not known~\cite{kraska2018case,ding2020alex,nathan2020learning}.

\textbf{Range indexes} An alternative to on-the-fly binary search over sorted data is to keep the data in an index structure. Nonetheless, indexes that are built to answer range queries (such as B-trees) are similar to the binary search in that they need to keep the data sorted internally. Common index structures for range index include skiplists, B+trees, and radix-trees. The B+-tree is cache-efficient, but requires pointer chasing, which incurs multiple cache misses~\cite{graefe2011modern}. There has been a tremendous effort to make binary search trees and B+-trees efficient on modern hardware. For example, FAST~\cite{kim2010fast} organizes tree elements efficiently to exploit modern hardware features such as the cache line and SIMD. Another common solution is to use compression techniques on the indexed keys, most notably as a radix-tree. Modern radix trees exploit hardware-efficient heuristics for fitting a distribution in memory (usually by building a heuristically-optimized compressed trie), such as adaptive radix index (ART)~\cite{leis2013adaptive,binna2018hot}, and Succinct Range Filter (SuRF)~\cite{zhang2018surf}. Skiplist is specifically efficient for concurrent updates workloads~\cite{xie2017parallelizing,sprenger2016cache}.%

\textbf{Learned index structures}
Learned range indexes~\cite{kraska2018case,neumann2008smooth,galakatos2019fiting,ding2020alex,llaveshi2019accelerating} have recently been suggested as an alternative to range indexes. In this approach, a model is trained from the data with the intent of capturing the data distribution and processing the queries more efficiently. %
We refer to the paper by Kraska et al.~\cite{kraska2018case}, which introduced the idea of the learned index. In a learned index, the CDF of the key distribution is learned by fitting a model, and the learned model is subsequently used as a replacement of the  index (B+-trees or similar) for finding the location of the query results on the storage medium. Index learning frameworks such as the RMI model~\cite{kraska2018case,marcus2020cdfshop} can learn arbitrary models~\cite{marcus2020cdfshop}, although a further theoretical study~\cite{ferragina2020learned} as well as a recent experimental benchmark~\cite{kipf2019sosd} have shown that simple model like linear splines are very effective for datasets. Spline-based learned indexes include Piecewise Geometric Model index (PGM-index)~\cite{ferragina2020pgm}, Fiting-tree~\cite{galakatos2019fiting}, Model-Assisted B-tree (MAB-tree)~\cite{hadian2020madex}, Radix-Spline~\cite{kipf2020radixspline}, Interpolation-friendly B-tree (IF-Btree)~\cite{hadian2019interpolation} and some others
~\cite{llaveshi2019accelerating,setiawan2020function}. We refer to \cite{ferragina2020survey} for an extensive comparison of learned indexes.
Recently, there has been numerous theoretical works~\cite{bilgram2019cost,sablayrolles2018spreading,sablayrolles2018d,li2019scalable} on learned indexes. Also, numerous efforts have been made to handle practical challenges around using a learned index, including update-handling~\cite{ding2020alex,hadian2019considerations} and designing a learned DBMS~\cite{kraska2019sagedb}. The idea of using a model of the data to boost an existing algorithmic index has been the center of focus in the past few years~\cite{graefe2011modern,hadian2019considerations,qu2019hybrid,hadian2020madex}. 
In the multivariate area, learning from a sample workload has also shown interesting results \cite{nathan2020learning,li2020lisa,hadian2020handsoff,dolatshah2015balltree}.
Aside from the main trend in learned indexes, which is on range indexing, machine learning has also inspired other indexing and retrieval tasks. This includes bloom filters~\cite{mitzenmacher2018model,dai2019adaptive}, inverted indexes~\cite{xiang2018pavo},  %
computing list intersections~\cite{ao2011efficient}, and multidimensional indexing on datasets with correlated attributes~\cite{hadian2021coax}.

\section{Conclusion and Future work}
Learning and modeling data distributions via machine learning approaches is a great idea for managing and analyzing data management systems. However, the approaches and objective functions that are common in machine learning problems are not necessarily optimal choices when the ultimate target is performance improvement. Instead of pushing machine learning model algorithm to its limits for highly accurate modeling of data distributions, it is more efficient if we only use ML models to approximate the high-level, generalizable "patterns" in data distribution (the holistic shape), and handle the fluctuations and fine-grained details of the distribution using a more hardware-efficient approach, outperforms learned models as well as algorithmic index structures even if a simple or somewhat dummy model such as min/max linear interpolation is used. The Shift-Table layer is effective in learning almost all distributions even without using models that require training from data, and takes only a single pass over the data points to build the layer. %
 Our results show that even a simple linear model equipped with the Shift-Table enhancement layer outperforms trained and tuned learned indexes by 1.5X to 2X on real-world datasets. 
 
 Our current work only considers read-only workloads. We leave it as future work to adapt Shift-Table with workloads having updates. One idea is to capture the drifts in data distribution using update-tracking segments~\cite{hadian2019considerations}, and use Fenwick trees to estimate and correct the drifts in both the model and the Shift-Table.
 
 \balance


\end{document}